\documentclass[aps,prl,10pt,twocolumn,groupedaddress,notitlepage,showpacs,floatfix]{revtex4-1}
\pdfoutput=1
\usepackage{graphicx,graphics,epsfig,subfigure,times,bm,bbm,amssymb,amsmath,amsthm,mathrsfs,MnSymbol}
\usepackage{gensymb}
\usepackage{amsfonts}
\usepackage[matrix,frame,arrow]{xypic}
\usepackage[pdfstartview=FitH]{hyperref}
\usepackage{subfigure}
\hypersetup{
    colorlinks=true,       
    linkcolor=red,          
   citecolor=magenta,        
    filecolor=magenta,      
    urlcolor=cyan,           
    runcolor=cyan
}
\usepackage[pdftex]{color}
\usepackage{braket}
\usepackage{enumerate}
\usepackage[normalem]{ulem}
\usepackage[usenames,dvipsnames]{xcolor}
\usepackage{multirow}
\usepackage{mathtools}

\definecolor{orange}{rgb}{1,0.5,0}

\newcommand{\ignore}[1]{}
\usepackage{geometry}
\newcommand{\red}{\color{red}}

\ignore{
\documentclass[eprintnumbers,amsmath,amssymb,onecolumn,a4paper,caption ]{article}
\usepackage{amsfonts}
\usepackage{amssymb}
\usepackage{mathrsfs}
\usepackage{mathbbold}
\usepackage{bbm}
\usepackage{mathrsfs}
\usepackage{dcolumn}
\usepackage{bm}
\usepackage{times,epsfig,amssymb,amsmath}
\usepackage{float}
\usepackage{subfigure}
\usepackage{geometry}\geometry{left=2.5cm,right=2.5cm,top=3cm,bottom=3cm}
\newcommand{\red}{\color{red}}

\usepackage{color}

}

\begin{document}

\title{Thermalization of Topological Entropy after a Quantum Quench}
\author{Yu Zeng$^1$, Alioscia Hamma$^2$\footnote{Corresponding author.\\email address: ahamma@mail.tsinghua.edu.cn}, Heng Fan$^1$}
        \affiliation{%
$^1$Institute of Physics, Chinese Academy of Sciences, Beijing 100190, China\\
$^2$IIIS, Center for Quantum Information, Tsinghua University, Beijing 100084, China}

\date{\today}
\begin{abstract}
In two spatial dimensions, topological order is robust for static deformations at zero temperature, while it is fragile at any finite temperature. How robust is topological order after a quantum quench? In this paper we show that topological order thermalizes under the unitary evolution after a quantum quench. If the quench preserves gauge symmetry, there is a residual topological entropy exactly like in the finite temperature case. 
 We obtain this result  by studying the time evolution of the topological 2-R\'enyi entropy in a fully analytical, exact way. These techniques  can be then applied to systems with strong disorder to show whether a many-body localization phenomenon appears in topologically ordered systems.
\end{abstract}
\maketitle

\section{Introduction}
The development of quantum many-body physics in recent years has opened the doors - both theoretically and experimentally - to the exploration of new quantum phases of the matter\cite{Sachdev-QPT, wenbook} and the behaviour away from equilibrium of quantum systems with many particles\cite{coldatoms1, coldatoms2, coldatoms3,PolkovnikovRMP}.

Novel quantum phases of matter that feature quantum or Topological Order (TO) cannot be described by the usual theory of symmetry and symmetry breaking, and therefore are not characterised by a local order parameter\cite{wen:1990, wenbook,goldenfeld, nussinov, correlations2}. They possess topological degrees of freedom, and excitations described by a topological quantum field theory\cite{freedman:2002}. Moreover, they possess a long-range pattern of entanglement dubbed Topological Entropy (TE) that serves as non-local order parameter for these phases\cite{hamma:2005a, levin:2006, kitaevpreskill, hamma:2005b, hiz3, Hamma2008prb, isakov:2011, trebst:2007, ikim, papa:2007, negativity1, negativity2, grover}.  The topological entropy is associated to both the existence of a robust qubit\cite{hammamazac, dennis, bt} and of anyonic excitations. These topological characteristics  make these states  robust against a model noise based on local interactions, as it is very reasonable for the environment. For this reason, they are believed to be of great advantage for the implementation of quantum information processing, a paradigm dubbed as Topological Quantum Computing (TQC)\cite{tqc, kitaev:2003,nayak:2008, freedman:2002, tqc2}.

On the other hand,  coherent quantum dynamics has recently  become accessible to experimental inquiry and study, in systems realized by means of ultra cold atom gases in optical lattices\cite{coldatoms1, coldatoms2, coldatoms3}. The flexibility in engineering interactions in optical lattices makes them a very interesting way to implement Hamiltonians featuring Topological Order. The dynamics is obtained through the protocol of Quantum Quench \cite{quenchbook}. The Hamiltonian of the system $H(\lambda)$ depends { smoothly} on a set of external parameters $\lambda$ that are easy to control, like some coupling strength or strength of external fields.  The system is initially prepared in the ground state of $H(\lambda_0)$ for some value $\lambda_0$ of the external parameters, that at time $t=0$ are suddenly switched to the quench value $\lambda_1$. The initial state will then evolve unitarily by means of the evolution generated by $H(\lambda_1)$\cite{quench1,quench2,quench3,armin, dyn2}.

The main goal of this paper is to show the fate of topological order in  the Kitaev's toric code\cite{kitaev:2003}  following a quantum quench, by looking at the time evolution of the TE. The motivation for this study is three-fold. First of all, if TO should be robust against perturbations in the Hamiltonian, this has to hold also for time varying stray magnetic fields, which would place the system away from equilibrium. The second motivation comes from the general relationship between quench dynamics and thermalization in a closed quantum system. In recent years, there has been a flourishing of results (see, e.g., \cite{olshani, eth1,eth2, popescu, gogolin,linden:2009} and \cite{PolkovnikovRMP} for extended references) about the foundations of quantum statistical mechanics. It has been understood that, if dynamics is complex enough - which is a generic situation - then a closed quantum system will make local observables thermalize, as the rest of the system can act as a thermal bath for the subsystem, although everything is away from equilibrium. One thus wonders whether topological degrees of freedom need to thermalize as well.  Finally, this kind of study gives us a handle to deal with the problem of finite temperature from a different point of view. If one can show that a system is robust against a quantum quench, then one can hope that the system may display robustness also at finite temperature.

 The main result of the paper, is that TO after a quantum quench will equilibrate to the thermal value at finite temperature\cite{castelnovo:2006, finiteT, finiteT2,chamon3d}.
. That is, zero if no symmetry is imposed, or half of the initial value if the gauge structure is conserved.
 Therefore, in 2D, TO in the toric code will not survive a quantum quench.

This result implies that the time evolution is quite different from a static perturbation. As it has been shown in a number of papers, the topological phase is robust against local perturbations of the Hamiltonian\cite{trebst:2007,Hamma2008prb,castelnovo:2008,halasz:2012a, jahromi:2013, dusuel2}. The study of static perturbations, that is, the effect that a perturbation has on the new ground state, has recently been amenable to numerical study on large systems using two dimensional DMRG methods\cite{cincio:2013, hamma:2013}. It has been found that regardless of how the system is perturbed, the topological phase is robust for some finite strength of the perturbation\cite{hamma:2008, trebst:2007, jvidal:2009, jvidal:2009b, wu:2012, karimipour:2013, dusuel2}. This means that { not only the phase is robust until a critical value is reached} - this has also been proven analytically in some remarkable papers\cite{klich, spyros1, spyros2}, but also that all the topological features are robust. They are indeed, properties of the phase. In particular, it has been shown that the TE is robust.

However, as we pointed out, the dynamics after a quantum quench has an entirely different physics. If the system dynamics is complex enough, every local subsystem can see the rest of the system as an environment, and thermalize locally, in spite of the fact that the global evolution is unitary.  In this case, topological order in 2D would be destroyed. Nevertheless, local thermalization does not necessarily imply whether topological degrees of freedom or { topological} observables do indeed thermalize. { After all, one of the goals of using topological states of the matter for quantum computation is to have some topological observables that do not decohere or thermalize while typical local observables will.} So one cannot just borrow this picture and draw conclusions. One needs to calculate. This issue has been explored in the past by some of us\cite{halasz:2012b}, and other authors\cite{kay,otherquench, armin}. The complexity of dealing with the time evolution of a quantum system, is, of course, formidable. Numerical analysis is limited to very small system sizes. In \cite{tso} was indeed found that TO and TE would not survive certain quenches that would break the gauge symmetry, but because of very small system sizes the results were not conclusive. From the analytical side, previous results relied on the simplification stemming from the restriction to a gauge-preserving quench\cite{halasz:2012b}. In this case, it was shown that the topological entropy computed from a  subsystem with spins only on the boundary but no bulk, which we call thin subsystem, in the 2D toric code is robust. In\cite{armin}, the effect of the breaking of integrability was shown to be unable to create topological order, together with a volume law for entanglement, thus suggesting that quantum quenches would be like thermalization for the toric code.

 In this paper, we attack the problem of a quantum quench of the 2D toric code without requiring that it preserves any symmetry of the system, including the gauge structure. We present a fully analytical solution of the problem, by developing and extending to the time domain a technique presented in \cite{santra2014}.  The technique also allows us to study the TE associated with a subsystem with a bulk instead of a thin subsystem. We  show that the presence of the bulk is very important and that in the time evolution the difference between thin and thick subsystem is critical, whereas in the case of static perturbations both subsystems yield similar results. In order to perform a fully analytical treatment, we choose the quench so that the system is completely integrable in free fermions, dubbed the $\tau$ picture \cite{spinmapping1,spinmapping2}.  To be sure, one would doubt that if the evolving Hamiltonian is fully integrable, there could be no thermalization at all. Indeed, as a whole the system does not thermalize. There is an extensive number of conserved quantities in the so called $\tau$ picture. However, that does not mean that no observable would not thermalize. Here, we show that the TE will thermalize. In 2D, topological order is even more fragile than local degrees of freedom. Again, we highlight that using an integrable quench is due to finding a full analytic solution. A {\em fortiori} though, under a non-integrable quench, no degree of freedom, topological or not, will be conserved.
In order to apply these techniques in the non integrable case, one could resort to perturbation theory following the lines of \cite{dusuel:2011} and \cite{halasz:2012b}. Moreover, one can pair the same technique with numerical techniques to study systems with disorder.



\section{Topological R\'enyi Entropy after a gauge-breaking quantum quench}
\begin{figure}
\centering
\includegraphics[width=0.47\textwidth]{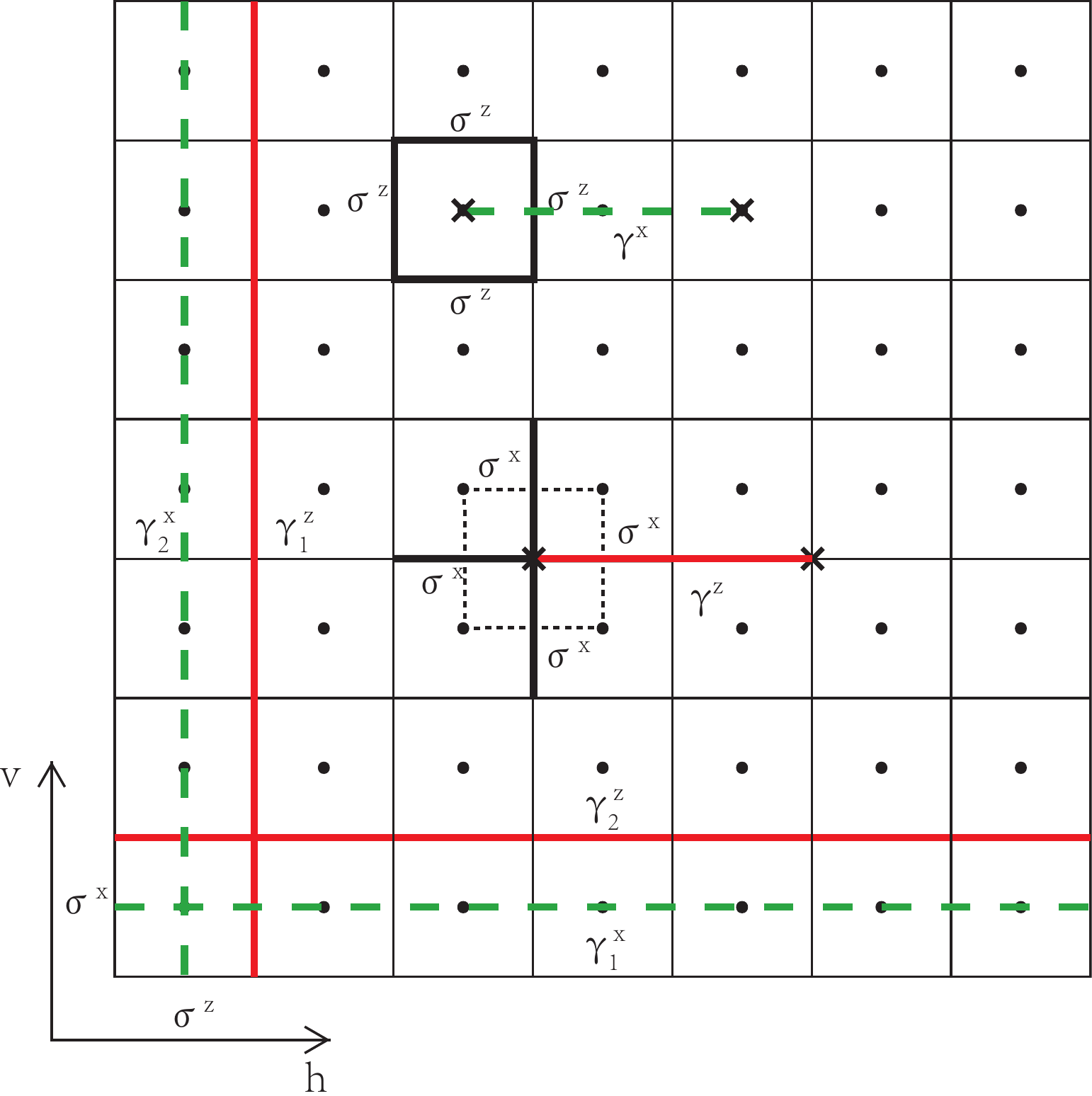}\\
\caption{(color online) A $N\times N$ square lattice with periodic boundary condition. It showed the star operator and the plaquette operator. there are two types of open strings corresponding to two type of excitations. It also showed the horizontal edge (along the direction h) and the vertical edge (along the direction v). We arrange fields in the $+z$ direction with magnitude of $\lambda_z$ on the horizontal edges and fields in the $+x$ direction with $\lambda_x$ on the vertical edges.}\label{lattice}
\end{figure}

We start with the toric code model introduced by Kitaev\cite{kitaev:2003}. The Hamiltonian for this model defined on a square lattice with $N\times N$ sites with spins $1/2$ on the bonds is given by
 \begin{eqnarray}\label{HTC}
 H_{TC}=-\sum_{s}A_{s}-\sum_{p}B_{p}
 \end{eqnarray}
where the star operators $A_{s}\equiv\prod_{i\in s}\sigma_{i}^{x}$ and the plaquette operators $B_{p}\equiv\prod_{i\in p}\sigma_{i}^{z}$ belong to stars(s) and plaquettes (p) on the lattice containing four spins each, see Fig.\ref{lattice} . This model features topological order in the ground state. If we add a perturbation $V(\lambda)$ that is the sum of local operators, for a finite range of $\lambda$ topological order is preserved\cite{hamma:2008, trebst:2007,Hamma2008prb,castelnovo:2008,halasz:2012a, jvidal:2009, jvidal:2009b, wu:2012, karimipour:2013}. On the other hand, if the system is put in contact with a heat reservoir and we wait for thermalization to happen (or if we do prepare the system in the Gibbs state), the topological order is destroyed.
How do we detect topological order in a system? It is very remarkable that topological order is detected and consists in a particular pattern of entanglement in the wave-function.
Entanglement is the most defining property of quantum mechanics. If something is genuinely quantum, that is, it cannot be simulated or explained with just classical concepts, one needs to take in account entanglement\cite{amico, nielchuang}. For this reason, quantum order or quantum phases of matter must have some non trivial pattern of entanglement\cite{kitaevpreskill, levin:2006}, and a big part of the recent effort in condensed matter theory and quantum field theory resides in the calculation of entanglement. However, as measured by the von Neumann entropy, entanglement is a formidable quantity to compute and measure. It requires the knowledge of all the eigenvalues of the reduced density matrix. So it requires perfect knowledge of the wave-function, which is a very hard task from the analytical, numerical and experimental point of view. If one wants to use entanglement properties as an order parameter, one should look for quantities that are in principle measurable, that is, they are the expectation value of some hermitian operator, and the hermitian operator must not explicitly depend on the wave-function itself. What choices do we have? We can consider the generalization of the von-Neumann entropy  to a family of entropies known as $\alpha-$R\'enyi entropies, defined as
\begin{eqnarray}
S^{AB}_{\alpha}\equiv\frac{1}{1-\alpha}\log_{2}Tr[\rho^{\alpha}_{A}]
\end{eqnarray}
associated to the reduced density matrix $\rho_A =\mbox{Tr}_B \rho$ after a tensor product structure of the Hilbert space  $\mathscr{H} = \mathscr{H}_A \otimes \mathscr{H}_B$.
In the case of  $\alpha=2$, though, we can find a very useful interpretation of the R\'enyi entropy:
\begin{eqnarray}\label{S2}
S^{AB}_{2}=-\log_{2}Tr[\rho^{2}_{A}]=-\log_2 P.
\end{eqnarray}
where $P$ is the purity of the state $\rho_A$. { Now, this quantity is a simple function of an observable. We need to first prepare two copies of $\rho\rightarrow\rho\otimes\rho\in  (\mathscr{H}_A \otimes \mathscr{H}_B)\otimes ( \mathscr{H}'_A \otimes \mathscr{H}'_B)$. Then, considering the two copies of   $\rho_A$, namely $\rho_A^{\otimes 2}$, one has}
\begin{eqnarray}
P = Tr[\rho^{2}_{A}] = Tr[\mathbb{S}_A \rho^{\otimes 2}]
\end{eqnarray}

\begin{figure*}
\centering
\includegraphics[width=1\textwidth]{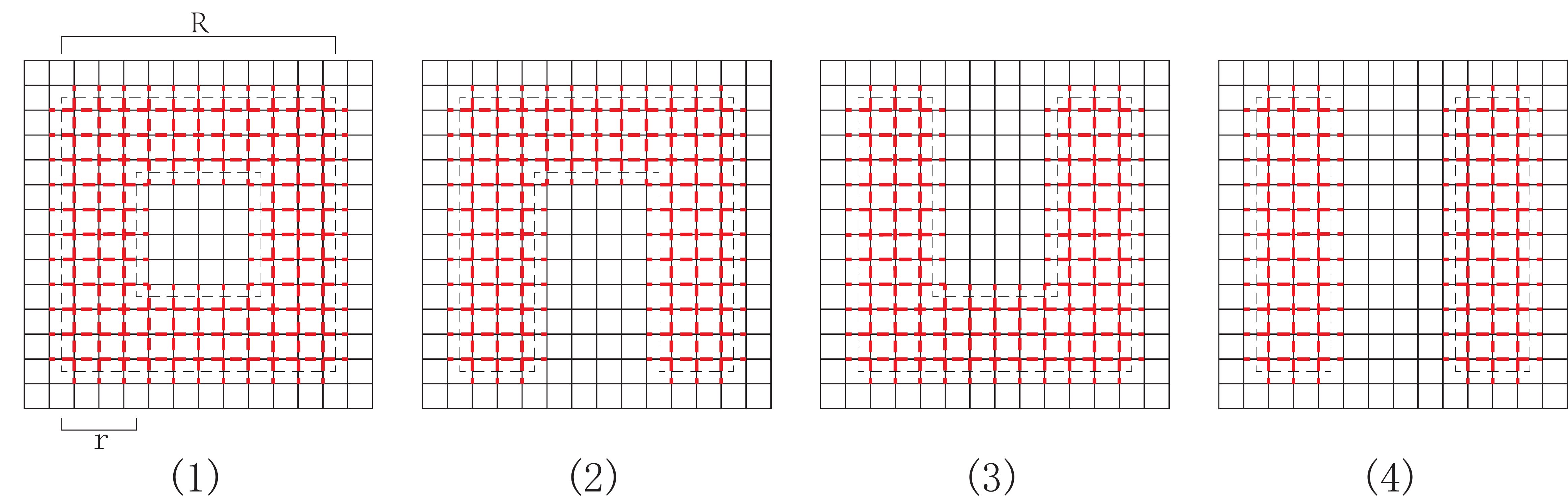}\\
\caption{(color online) Illustration of the subsystems A (red dashed lines) and B (black lines) in the four cases that are applied to calculate the topological entropy with extension R=11, thickness r=3 .}\label{4plots}
\end{figure*}
where $\mathbb{S}_A$ is the swap operator between the two copies of $\rho_A$\cite{swap}. Here is a simple proof. An arbitrary state $\rho$ in $\mathscr{H}$ can be written as $\rho=\sum_{i_Ai_Bj_Aj_B}\alpha^{i_Ai_B}{}_{j_Aj_B}|i_A,i_B\rangle\langle j_A,j_B|$, where $|i_A\rangle$ and $|i_B\rangle$ are the bases in $\mathscr{H}_A$ and $\mathscr{H}_B$ respectively and the coefficient $\alpha^{i_Ai_B}{}_{j_Aj_B}$ satisfies hermitian constrain $\alpha^{i_Ai_B}{}_{j_Aj_B}^\ast=\alpha^{j_Aj_B}{}_{i_Ai_B}$. Actually $\rho$ is a tensor of type (2,2), and we use Einstein summation convention for simplicity, namely, $\rho=\alpha^{i_Ai_B}{}_{j_Aj_B}|i_A,i_B\rangle\langle j_A,j_B|$.  then $\rho^{\otimes 2}=\alpha^{i_Ai_B}{}_{j_Aj_B}\alpha^{k_Ak_B}{}_{l_Al_B}|i_A,i_B\rangle\langle j_A,j_B|\otimes|k_A,k_B\rangle\langle l_A,l_B|$. Note that $\mathbb{S}_A$ `swaps' the two copies's component in $\mathscr{H}_A$ and $\mathscr{H}^\prime_A$ as $\mathbb{S}_A|i_A,i_B\rangle\langle j_A,j_B|\otimes|k_A,k_B\rangle\langle l_A,l_B|=|k_A,i_B\rangle\langle j_A,j_B|\otimes|i_A,k_B\rangle\langle l_A,l_B|$. In the end we calculate the trace and finally get $Tr[\mathbb{S}_A \rho^{\otimes 2}]=\alpha^{i_Ai_B}{}_{k_Ai_B}\alpha^{k_Ak_B}{}_{i_Ak_B}$. We can see that the swap operator actually `swaps' the contraction indexes of the tensor. It is easy to verify that the result is the same when we calculate $Tr[\rho^{2}_{A}]$.  So the $2-$R\'enyi entropy simply reads $S^{AB}_{2}=-\log_2 P=-\log_2Tr[\mathbb{S}_A \rho^{\otimes 2}]$.
Now,  the von Neumann entropy is the unique measure of bipartite entanglement that quantifies\cite{vedralnature, amico} the conversion into Bell pairs that one can obtain from a state, which is important for quantum information processing protocols. However, as far as properties of the phase are concerned in condensed matter, the R\'enyi entropy is as good. It marks quantum phase transitions in the same way, and, when entanglement characterises a phase, $S^{AB}_{2}$ does it as well. In particular, it has been shown\cite{flammia:2009} that TE measured by the R\'enyi entropy detects TO in exactly the same way.
The fact that the $2-$R\'enyi entropy is the expectation value of an observable also makes it possible to conceive realistic scenarios for its measurement\cite{Zoller2012, Abanin2012, Ekert2002}. This is important as the search of quantities that can be measured for detecting topological order is one of the most important topics in the field. So the topological part of the $2-$R\'enyi entropy is a possible candidate, together with other measures, see\cite{spectroscopy, correlations}.
Following\cite{kitaevpreskill, levin:2006}, the topological R\'enyi entropy is defined as the linear combination of four  R\'enyi entropies associated to four different regions $(1), (2), (3), (4)$, see Fig.($\ref{4plots}$):
\begin{eqnarray}\label{ST}
S^{T}_{\alpha}\equiv -S^{(1)}_{\alpha}+S^{(2)}_{\alpha}+S^{(3)}_{\alpha}-S^{(4)}_{\alpha},
\end{eqnarray}

For the toric code model $H_{TC}$, the topological entropy reads $S^T_\alpha=2$, for every $\alpha$\cite{flammia:2009}. As mentioned above, for a finite range of $\lambda$, { the system stays in the topological phase, and accordingly, }the value of $S^{T}_2(\lambda)$ in the new ground state of the Hamiltonian
 \begin{eqnarray}\label{perturbed}
 H(\lambda)=-\sum_{s}A_{s}-\sum_{p}B_{p} +V(\lambda)
 \end{eqnarray}
is preserved, in the limit of $R,r\rightarrow\infty$\cite{trebst:2007,Hamma2008prb,castelnovo:2008,halasz:2012a}. On the other hand, for every finite temperature $\beta$, the value of $S^T$ in the thermal state $\rho= Z^{-1}e^{-\beta H_{TC}}$ goes to zero in the thermodynamic limit. However, if one freezes one of the series of quantum numbers $A_s$ or $B_p$ (i.e., if one enforces a gauge symmetry), the value of $S^T$ goes to one half of the full value, namely $S^T(gauge) =1$, signalling a classical form of topological order in the system\cite{finiteT,castelnovo:2006,chamon3d}.

In this paper, we want to understand the dynamics of $S^T$ after a quantum quench. The protocol of the quantum quench is simple. We prepare the system in the state $\Psi(0)$ being in the ground state of $H_{TC}$, and then, at $t=0$, we suddenly switch on the term $V(\lambda)$ in $H(\lambda)$. The wave function of the system will then evolve unitarily as
\begin{eqnarray}
|\Psi(t)\rangle = e^{-i H(\lambda) t}|\Psi(0)\rangle.
\end{eqnarray}
Taking the trace over the degrees of freedom in $B$ of the above state, we can obtain the time evolution $\rho_A(t)$ of the partial state. We set on studying the presence of topological order after a quantum quench by studying then the quantity $S^T_2(t) = S^T_2(\rho_A(t))$.  In order to find a fully analytical solution of this problem, we need to find a perturbation $V(\lambda)$ of the toric code such that the model is still completely integrable. In this way, one can study exactly both the ground state manifold and the time evolution after a quantum quench as the perturbation $V(\lambda)$ is switched on\cite{halasz:2012a,halasz:2012b}. { Again, we want to highlight that using an integrable model is the right thing to do if one wants to prove fragility. If we found that TE is robust under an integrable quench, we could suspect that thermalization fails to happen just because of the many conserved quantities. But if TE thermalizes under an integrable quench, it will even more so do if the evolving Hamiltonian is non integrable. }
In particular, let us see how the toric code with a certain arrangement of the external fields can be mapped into a system of free fermions\cite{santra2014, pfeuty:1970}. We write $V(\lambda)$ as the sum of  external fields on the bonds, with  a special arrangement: the field in the $+z$ direction with magnitude of $\lambda_{z}$ on the horizontal(h) edges and the field in the $+x$ direction with $\lambda_{x}$ on the vertical(v) edges. In this model, there are $2N^{2}$spins on the edges of $N\times N$ square lattice with periodic boundary conditions. The Hamiltonian for the model then reads:
 \begin{eqnarray}\label{Hsigma}
 H(\lambda)=-\sum_{s}A_{s}-\sum_{p}B_{p}-\lambda_{z}\sum_{i\in h}\sigma_{i}^{z}-\lambda_{x}\sum_{i\in v}\sigma_{i}^{x},
 \end{eqnarray}
Notice that when one of the $\lambda_x, \lambda_z$ is zero, the system { preserves} one of the two local $\mathbb Z_2$ gauge symmetries $[H(\lambda_z),B_p]=[H(\lambda_x),A_s]=0$ for every $s,p$. One of the main goals of this paper is to find results when no gauge symmetry is imposed on the system, namely $\lambda_x\ne0, \lambda_z\ne 0$.
After exact diagonalization, one can obtain an analytic expression for $|\Psi(t)\rangle$. Moreover, one can calculate analytically all the many-spin correlation functions as a function of time. This is a key point to obtain what we want. Indeed, from the technical point of view, the main result is that we can compute  the topological $2-$R\'enyi entropy  as a function of time $t$, and quench strengths  $\lambda = (\lambda_x, \lambda_z)$for the time evolution after a quantum quench, namely
\begin{eqnarray}\label{qdyn}
 S^T_2(t, \lambda)=log_2\left(\frac{P^{(1)}(t, \lambda)P^{(4)}(t, \lambda)}{P^{(2)}(t, \lambda)P^{(3)}(t, \lambda)}\right)
\end{eqnarray}
where $P(t, \lambda)$ is the purity of the evolved subsystem $A$, namely
\begin{eqnarray}\label{ppp}
P(t) = \mbox{Tr}_A  \left[ \mbox{Tr}_B\left( e^{-iH(\lambda)t }  |\Psi(0)\rangle\langle\Psi(0)| e^{iH(\lambda)t } \right)\right]^2.\nonumber\\
\end{eqnarray}
The mapping into free fermions mentioned above proceeds from a first mapping of the physical spins on the links, which we call '$\sigma$-picture', to some effective spin on the sites, of both the initial lattice and the dual lattice (i.e., the sites of the plaquettes). This picture is called here the `$\tau$-picture', H($\lambda$). This mapping brings the model $H(\lambda)$ into the  sum of the Ising chains in transverse field over $2N$ different lines, namely, $N$ rows on the lattice, and $N$ rows on the dual lattice.  The  eigenspace of H($\lambda$) in `$\tau$-picture' is a  tensor product over the different chains. At this point, we show that $P(t)$ can be calculated exactly by sum of  correlation functions\cite{halasz:2012b,santra2014, santra2014, Lieb1961, barouch:1971}. The details of the mapping and the solution in terms of correlation functions are shown in the Appendix.
\begin{figure*}[!htb]
\begin{center}
\includegraphics[width=1\textwidth]{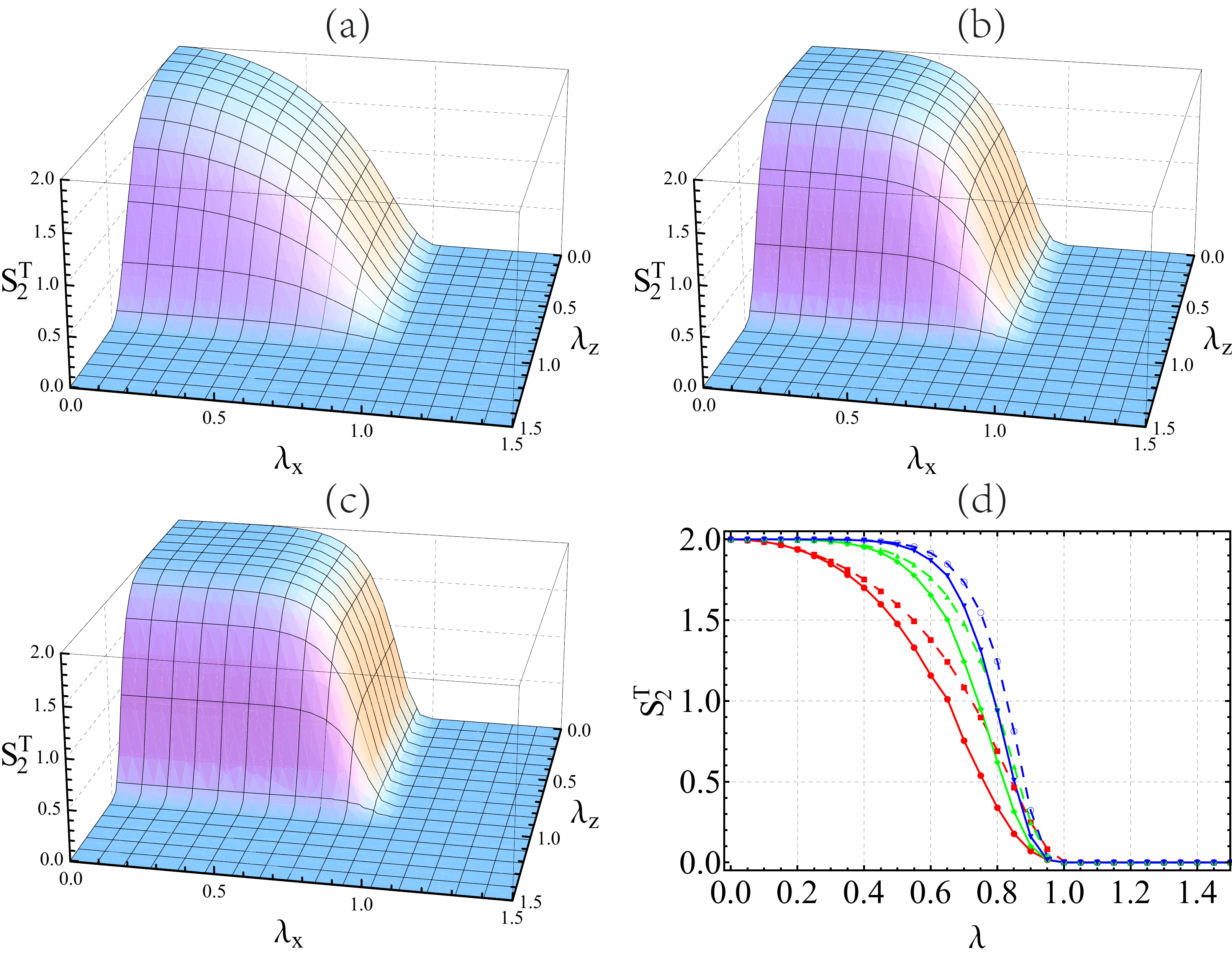}\\
\caption{(color online) Topological R\'enyi entropy $S_2^T$ in static case as a function of fields $\lambda_x$ and $\lambda_z$ with different system sizes. (a) R=5, r=1, N=100; (b) R=8, r=2, N=160; (c) R=11, r=3 , N=220. (d) Illustration of $S_2^T$ with Hamiltonian preserving $Z_2$ gauge symmetry ($\lambda_z=0, \lambda_x=\lambda$, dashed lines) and breaking gauge symmetry ($\lambda_z=\lambda_x=\lambda$, solid lines). Various colors represent distinct system sizes. Red: R=5, r=1, N=100; green: R=8, r=2, N=160; blue: R=11, r=3, N=220.}\label{static}
\end{center}
\end{figure*}
  The main result of this paper is obtaining a closed formula for the $2-$R\'enyi Topological Entropy, after a quantum quench. This is given by substituting the following expression for the purity of the state $\rho_A$ into Eq.(\ref{ppp})
\begin{widetext}
\begin{eqnarray}\label{purityt}
P(t) =C_P\sum_{\partial\tilde{g}\in \partial G^\prime_{A}} \sum_{\substack{\tilde{g}\in G^\prime_A\\ \tilde{z}\in Z_A}}|\langle\tilde{g}\partial\tilde{g}\tilde{z}\rangle_{\Psi(t)}|^2\sum_{\substack{\tilde{h}\in H^\prime_A\\ \tilde{x} \in X^{\prime}_{A}}}|\langle\tilde{x}\partial\tilde{x}(\partial\tilde{g})\tilde{h}\rangle_{\Phi(t)}|^2
\sum_{\partial\bar{g}\in \partial G^\prime_{B}}(-1)^{\partial\bar{g}\partial\bar{x}(\partial\bar{g})\cap\tilde{z}\tilde{h}}.
\end{eqnarray}
\end{widetext}
In the above formula, $\Psi(t) \otimes(\Phi(t))$ describes the time evolution of the system ($\psi$ and $\phi$ refer to the quantum numbers on two different sub lattices).  The operators $\tilde{g}, \partial\tilde{g}, \tilde{z}, \tilde{x}, \partial\tilde{x}, \tilde{h}$ and $\partial\bar{g}, \partial\bar{x}$ represent string operators operating with the Pauli algebra on the spins in the lattice, either in the subsystems $A,B$. The phase factor takes in account whether such operators commute or anti-commute. As we can see, the evaluation of this formula requires just the knowledge of correlation functions. As the system $H(\lambda)$ is integrable, all these quantities can be obtained analytically.
Notice that for $t=0$, this is the topological entropy in any given eigenstate of the system.
The derivation of Eq.(\ref{purityt}) is far from being trivial, and it requires several pages of calculations. The full derivation is presented in the Appendix, where it appears as Eq.(\ref{purity}).

An important remark regards system sizes. All the formulae above have been obtained in the thermodynamic limit for the size of the lattice, namely $N\rightarrow\infty$. They also hold for every size of the subsystem $R,r$. As one can see, though, the number of correlation functions to compute grows exponentially with the size of the subsystem $A$.  For example, for a subsystem of the type (1) with $R, r$ (see Fig.\ref{4plots}), the number of correlation functions to compute   scales as $2^{2R+2r}$. Computation of each correlation function is reduced to computation of a determinant whose maximum dimension is $R+1$. \cite{Lieb1961, barouch:1971} Although the number of correlation functions to compute is exponential, this calculation can be effectively carried over also for large $R,r$ by using parallelization on high performance computing clusters.
\section{Results and discussion}
\begin{figure*}
\begin{center}
\includegraphics[width=1\textwidth]{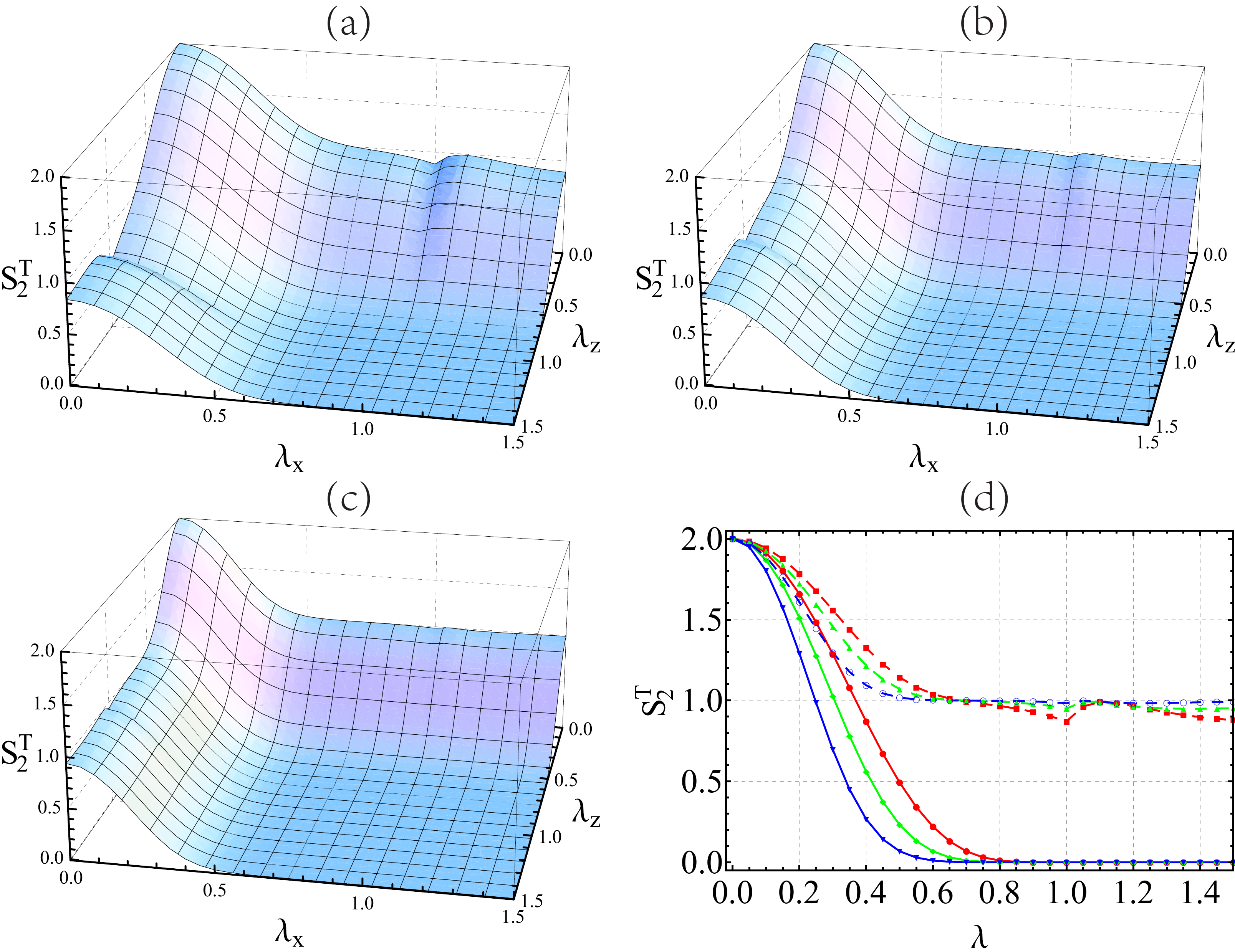}\\
\caption{(color online) Topological R\'enyi entropy $S_2^T$ in quantum quench case as a function of fields $\lambda_x$ and $\lambda_z$ with different subsystem sizes in thermodynamic limit ($N\rightarrow\infty$) and infinite time limit ($t\rightarrow\infty$). (a) R=5, r=1; (b) R=8, r=2; (c) R=11, r=3. (d) Illustration of $S_2^T$ with quench Hamiltonian preserving $Z_2$ gauge symmetry ($\lambda_z=0, \lambda_x=\lambda$, dashed lines) and breaking gauge symmetry ($\lambda_z=\lambda_x=\lambda$, solid lines). Various colors represent distinct subsystem sizes. Red: R=5, r=1; green: R=8, r=2; blue: R=11, r=3.}\label{tinftyevol}
\end{center}
\end{figure*}
\begin{figure*}[!htb]
\begin{center}
\includegraphics[width=1\textwidth]{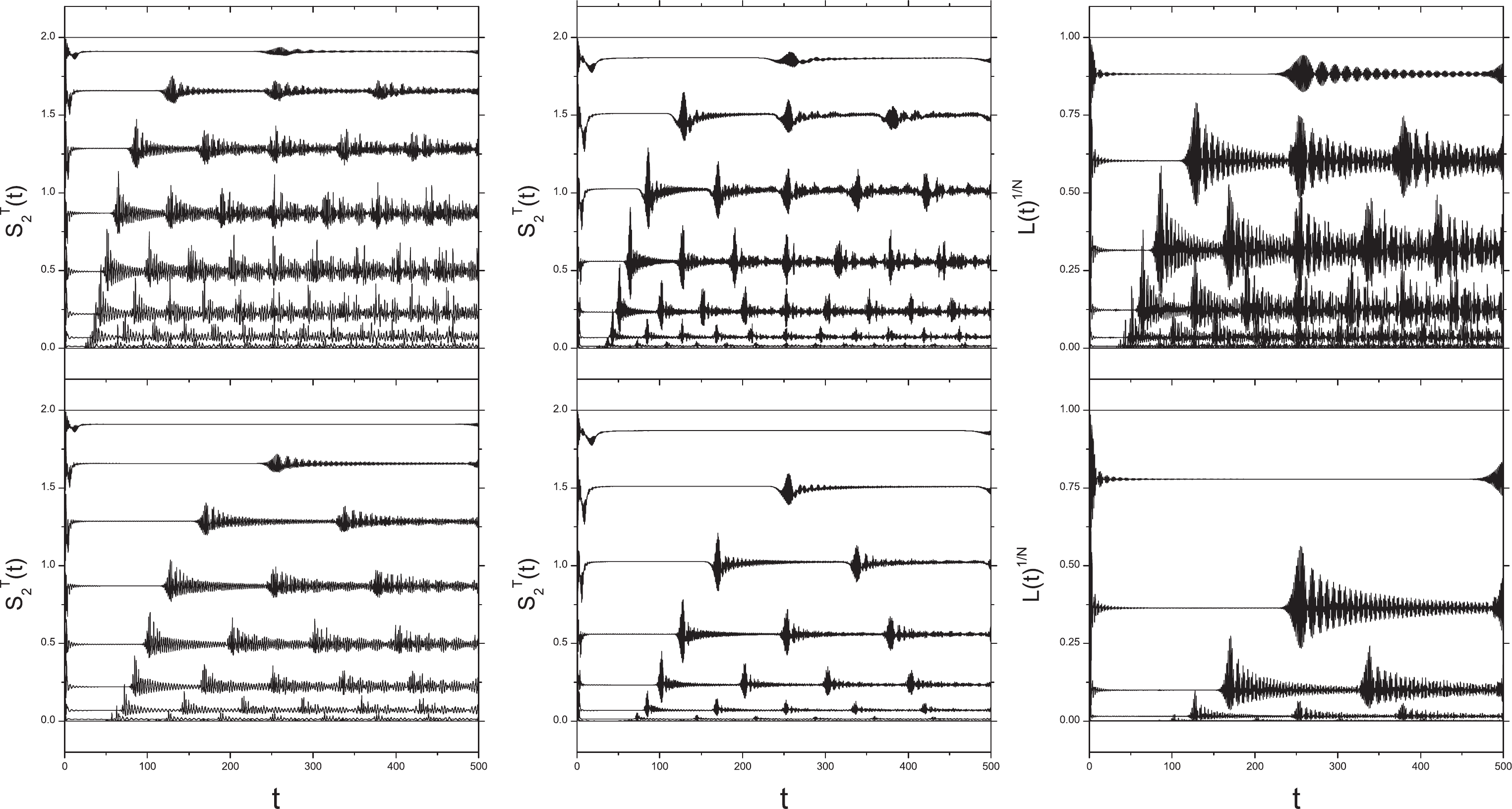}\\
\caption{ Illustration of (a)(b) Topological R\'enyi entropy and (c) Loschmidt echo after a quantum quench in finite time and with finite system size. In each plot, distinct 16 curves from top to bottom correspond to the different $\lambda=\lambda_x=\lambda_z$ from 0 to 1.5 in interval of 0.1. The subsystem and system size are (a1) R=5, r=1, N=100; (a2) R=5, r=1, N=200; (b1) R=8, r=2, N=100; (b2) R=8, r=2, N=200; (c1) N=100; (c2) N=200. We can see that the revival time of TE and LE is proportional to the system size N and is also dependent on $\lambda$, which means $t_\ast\sim N/v(\lambda)$. $v(\lambda)$ is the speed of signals in the system given by the LiebRobinson bound as $v\sim\lambda$. The time average value of TE for each $\lambda$ is equal to TE of the dephased state, which converge to the value for thermodynamic limit ($N\rightarrow\infty$) and $t\rightarrow\infty$ limit. We can also find that, as the size of the subsystem getting larger, TE is smaller with the same quench Hamiltonian $H(\lambda)$. After a quench, TE is almost completely destroyed when (a) $\lambda>0.8$ and (b) $\lambda>0.7$, which is compatible to the situation of $N\rightarrow\infty$ and $t\rightarrow\infty$.  }\label{tdep}
\end{center}
\end{figure*}
In this section, we show the results obtained from the computation of Eqs.(\ref{qdyn}, \ref{purityt}).

Let us first show the effect of a static perturbation. This amounts to compute Eq.(\ref{ppp}) in the instantaneous ground state ($t=0$) as $\lambda$ is varied . In Fig.\ref{static}
we can see the effect of a perturbation in the ground state. After the critical point $\lambda=1$, topological order is destroyed and TE vanishes. As the system size increases, the transition becomes sharper. As we can see, the presence of the gauge structure makes TO more resilient, and the TE vanishes in a smoother way. If gauge structure is destroyed, the transition is much sharper.
The result in presence of gauge symmetry is in complete according with earlier results \cite{halasz:2012a} both numerical and analytical.
Of course, all the effort carried so far was with the goal of computing time evolution, because that is the situation in which numerics will not help. Moreover, we are interested in a generic quench such that every symmetry (including the gauge symmetry) can be destroyed. This is indeed the case, as long as one of the $\lambda_x,\lambda_z$ is non-vanishing.

In Fig.\ref{tinftyevol}, we show the fate of TE measured by $S^T_2$ in the thermodynamic limit $N\rightarrow\infty$, at infinite times $t\rightarrow\infty$ for different subsystem sizes $R,r$. The result is fully analytical. We see that, as gauge symmetry is broken, the TE vanishes in the limit of large subsystem size. Some residual TE is alive for small subsystem sizes and moderate $\lambda$. On the other hand, if the gauge symmetry is preserved, (meaning either $\lambda_z, \lambda_x=0$), then $S^T_2=1$ for large subsystem sizes, which is half of the full value in the toric code. This is the main result of the paper: after a  quantum quench, at large times the system has the same topological entropy than in the thermal state. As it was shown in \cite{finiteT,castelnovo:2006, finiteT2, hammamazac}, if gauge symmetry is present, then the thermal state possesses a classical topological order with half  the value of the full topological entropy. This corresponds to the existence of a protected classical bit of information\cite{hammamazac}. On the other hand, if no gauge symmetry is preserved, in the thermodynamic limit all the TE disappears, corresponding to no possible information stored in a protected way in the system\cite{hammamazac}. Therefore, the main message is, {\em after a quantum quench, topological order thermalizes}. It is remarkable that this happens even when the quench is integrable. { This means that even though the system does not fully thermalize, as there are many conserved local quantities,  the evolution is complex enough to destroy the topological order and the topological observables of the system. } It is then quite natural to foresee the same scenario for a non-integrable quench, when there are not even local conserved quantities, and one would expect at large times to reach the Gibbs state locally.
In the next plot, Fig.\ref{tinftyevol}.d we show a one dimensional cross section of the above graph for clarity.

So far, we have presented the results for the limit of infinite time (and in the thermodynamic limit), as the correlation functions in Eq.(\ref{purityt}) have a compact analytic form in the limit $t\rightarrow\infty$ (see Appendix). However, we are interested also in understanding how fast thermalization is. We know that at infinite time, TE is completely destroyed (or halved with gauge symmetry) for the infinite subsystem, or it sets to a finite value for the finite subsystem if the quench parameter $\lambda$ is not too large. We will comment about the finite size effect of the subsystem at the end of this section. We ask ourselves at what characteristic time $t_{eq}$ this would happen. To this end, we just need to evaluate the time dependent correlation functions in Eq.(\ref{purityt})
The results are displayed in  Fig.\ref{tdep}, where we show the time evolution of $S^T_2(t)$ for a subsystem of size (a) $R=5,r=1$ and (b) $R=8,r=2$ for different values of the quench strength $\lambda=\lambda_x=\lambda_z$. We can clearly see that after a very short time TE thermalizes. Moreover, TE acquires some revivals at later times, well before the recurrence time (that is double exponential in the system size, see \cite{lorenz}). One can already see that the time scale for thermalization depends on the size of the subsystem $R,r$, although we do not have enough points to make an estimate. However, the structure of the revivals for different system sizes and strengths of the quench $\lambda$ is much clearer. Revivals are expected when the wave packet is partially reformed after signals in the system recombine\cite{lrh1, lrh2}. As the speed $v$ of signals in the system is given by the Lieb-Robinson bound as $v\sim\lambda$ \cite{lrh1, lrh2, liebrobinson0,liebrobinson1,liebrobinson2,liebrobinson3}, we can see that the time $t_*$ at which the revival on the profile of $S^T_2(t)$ is reached scales like $\lambda^{-1}$. Revivals in the full wave-function $|\Psi (t)\rangle$  are detected by the Loschmidt Echo (LE), defined as  ${L}_t :=|\langle\Psi (t)|\Psi (0)\rangle |^2$ \cite{LE}. We show the behaviour of LE in panel (c) of Fig.ref{tdep}. As one can see, TE and LE display the same time structure of revivals. Since $L_t$ is very difficult to detect for an extended system, as it rapidly (exponentially) shrinks to zero with system size\cite{LE}, it is actually desirable to find better observables to detect the structure of revivals. As we can see in Fig.ref{tdep}, the topological R'enyi entropy does detect the same revival times for a much larger set of system sizes, while the Loschmidt Echo is completely lost  for the system with $N=100$.
It is very interesting that one can use $S_2^T$ as a probe about both the thermalization and the witness of the system still being away from equilibrium. Thermalization under unitary evolution is in fact thermalization in probability, meaning that the probability of observing a value different from the typical value goes to zero in the thermodynamic limit. We  notice here, that in order to completely lose revivals, one also needs the thermodynamic limit of the subsystem, thus displaying the topological character of TE.

At this point, some more
 comments are in order regarding the subsystem size. In presence of small subsystem sizes, or, in the case the subsystem is 'thin", that is, consisting only of boundary, we find a residual topological entropy TE. However, the addition of a bulk makes it disappear (or reduce to its half value in case of gauge symmetry). In \cite{halasz:2012b} it was  showed that, for the thin subsystem, the full value $S^T_2=2$ was preserved. We find the same result with the formula presented here if we apply it to the thin subsystem. So the two results are in accord. Anyway, now we see that we have two ways of measuring $S^T$ that yield two different results. If $S^T$ is measured in a subsystem with bulk, we find thermal behaviour, while, on the other hand, if the subsystem is thin, we find a more robust behaviour. We are thus in a quandary, Which of the two ways is the right way to detect topological order? At zero temperature and for static perturbations the two ways give comparable results, but they have completely different behaviour in the dynamical picture. The two quantities must be associated to different aspects of topological order. As it was argued in \cite{halasz:2012b}, the topological entropy associated to a subsystem with bulk is associated to the existence of protected information in the system, and to the confinement-deconfinement transition for the topological quasiparticles\cite{confinement}. So we believe that this is the quantity of merit to detect topological order in a wave function. This opens up the question of what is the interpretation of the topological entropy for the thin subsystem, which will be investigated in the future.

\section{conclusions}
{
In this paper, we presented a fully analytical treatment of the time evolution of the Topological R\'enyi entropy $S^T_2$ after an integrable quantum quench. The main result of the paper is that $S^T_2$ quickly reaches a thermal value. Therefore, even though the quench is integrable, the dynamics is complex enough to make the topological order to thermalize. In two spatial dimensions, this amounts to destroying the topological order. One important consequence of this result, is that one can study dynamical or thermal stability of topological order in a unified way. It is thus conceivable that if topological order survives a quantum quench, then it would be also thermally stable, and viceversa. This opens the way to studying thermal stability of other models that feature topological order, like the toric code in higher dimensions\cite{dennis, hammamazac, toricboson}.  Moreover, the technique established here can be directly extended to the case of a  quench with strong disorder\cite{abaninlog, yuhammanew}. In one dimensional spin chains, strong disorder may cause many-body localization (MBL)\cite{abaninlog,mblpapers1,mblpapers2,mblpapers3,mblpapers4}. Very little is known about MBL in 2D and nothing about whether MBL is possible together with TO. In presence of disorder, the system would be not fully integrable, but still amenable of analytical treatment, as we can still map the system to free fermions, and then proceed numerically to diagonalize a $N\times N$ matrix \cite{isingbook, randomising}(as opposed to an exponentially large matrix). Moreover, one can use unitary perturbation theory\cite{dusuel:2011} in combination with our technique. In this way, we can explore directly if there is many-body localisation in presence of topological order\cite{pachos, pachos2}, and if localisation does protect it after a quantum quench, or in temperature.
}

\section{acknowledgments}
This work was supported in part by the National Basic Research Program of China Grant 2011CBA00300, 2011CBA00301 the National Natural Science Foundation of China Grant  61033001, 61361136003, 11574176 (A.H.), 91536108 (H.F.).

\section{APPENDIX}
\subsection{ Mapping to Ising chains}

The ground state manifold $\mathscr{L}$ of the TCM is 4-fold degenerate. Each ground state is the uniform superposition of closed strings. These closed strings can be arranged in four sectors according to contractible and non-contractible loops on the torus. The 4-dimensional algebra $L(\mathscr{L})$ is generated by  two pairs of topological operators ($W^{x}_{1},W^{z}_{1}$) and ($W^{x}_{2},W^{z}_{2}$). $W^\alpha_a$ is defined as
\begin{eqnarray}
W^\alpha_a=\prod_{j\in\gamma^\alpha_a}\sigma^\alpha_a ,\quad\alpha=x,z\quad a=1,2.
\end{eqnarray}
Each $\gamma^\alpha_a$ is a non-contractible curve along the toric on the lattice or the dual lattice, see Fig.\ref{lattice}. The external fields generate excitations described by open strings. Therefore,
when the fields are turned on, the ground state is a superposition of both closed and open strings.

\begin{figure}[t]
\centering
\includegraphics[width=0.4\textwidth]{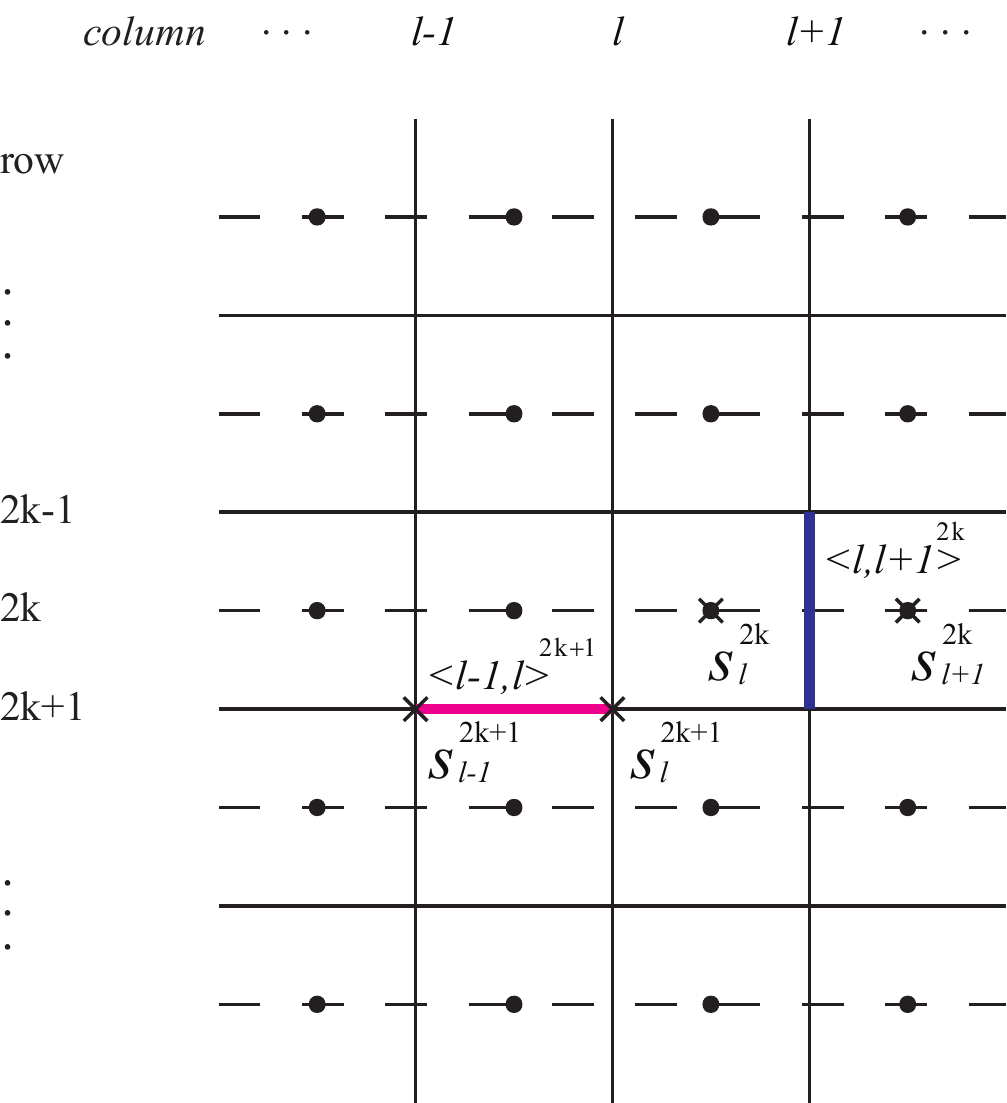}\\
\caption{(color online). Illustration of notations of site (cross) and links ( red and blue bold segments ) with row and column index. The physical spins live on the links ( '$\sigma$-picture')  while the effective spins live on the sites ('$\tau$-picture').  Site notation $s^i_j$ with odd (even) row index belongs to lattice (dual lattice).  }\label{lattice2}
\end{figure}

Because of the arrangement of the fields in horizontal and vertical lines, the Hamiltonian can be reduced
 into two mutually commutative part: $H=H_{1}+H_{2}$, where
\begin{eqnarray}
 H_{1}=-\sum_{s}A_{s}-\lambda_{z}\sum_{i\in h}\sigma_{i}^{z} \\
 H_{2}=-\sum_{p}B_{p}-\lambda_{x}\sum_{i\in v}\sigma_{i}^{x}.
\end{eqnarray}
 As it is easy to verify, the two  satisfy $[H_{1},H_{2}]=0$.
 For clarity, We use the symbol $s^{i}_{j}$ to denote the site of the lattice at row i and column j, and symbol $<j,j+1>^{i}$ to denote the bond located between $s^{i}_{j}$ and $s^{i}_{j+1}$, see  Fig.\ref{lattice2}.  Also, notice that $\{A_{s^{i}_{j}},\sigma^{z}_{<j,j+1>^{i}}\}=0$ and $\{A_{s^{i}_{j+1}},\sigma^{z}_{<j,j+1>^{i}}\}=0$. So we  can introduce the effective spins $\tau^{z}_{s^{i}_{j}}\equiv A_{s^{i}_{j}}$ and $\tau^{x}_{s^{i}_{j}}\equiv\prod_{k=1}^{j}\sigma^{z}_{<k-1,k>^i}$ ($\sigma^{z}_{<0,1>^i}=\sigma^{z}_{<N,1>^i}$ for periodic boundary condition) which satisfy $\{\tau^{z}_{s^{i}_{j}},\tau^{x}_{s^{i}_{j}}\}=0$ and commute with different site index, so $\sigma^{z}_{<j,j+1>^{i}}=\tau^{x}_{s^{i}_{j}}\tau^{x}_{s^{i}_{j+1}}$. $H_{1}$ can be mapped to the effective spin form:
 \begin{eqnarray}\label{h1}
\tilde{H}_{1}=-\sum_{i=1}^{N}\hat{O}_i\equiv -\sum_{i=1}^{N}(\sum_{j=1}^{N}\tau^{z}_{s^{i}_{j}}+\lambda_{z}\tau^{x}_{s^{i}_{j}}\tau^{x}_{s^{i}_{j+1}}).
\end{eqnarray}
Similary, $H_2$ will map into $\tilde{H}_{2}$.
Eq.(\ref{h1}) shows that each term $\hat{O}_i$ is an Ising chain. The chains are decoupled, and chains on different rows commute: $\left[ \hat{O}_l,\hat{O}_m\right]=0$. Terms in the bracket, which are all 1D Ising chains, are all mutual commutative with different row index,  so was $H_{2}$. We can add each Ising chains in $\tilde{H_{1}}$ and $\tilde{H_{2}}$ together. All we need to do is to extend the row numbers from N to 2N where $\tilde{H_1}$ contains all the lines with the odd row numbers while $\tilde{H_2}$ contains the even ones . In this condition the site indexes contain both lattice and dual lattice. So the original Hamiltonian can be mapped as an array of two different types of Ising chains. The $i=$odd chains are horizontal lines on the lattice, while the $i=$even ones are vertical lines on the dual lattice, see Fig.\ref{lattice2}. Thus we have
\begin{eqnarray}\label{isingH}
\tilde{H}&=&-\sum_{i=1}^{2N}  \hat{K}_i \equiv -\sum_{i=1}^{2N}\left(\sum_{j=1}^{N}\tau^{z}_{s^{i}_{j}}+\lambda(i)\tau^{x}_{s^{i}_{j}}\tau^{x}_{s^{i}_{j+1}}\right)\\
\centering\lambda(i)&=&\lambda_{z},\quad \text{i is odd;} \nonumber \\
\centering\lambda(i)&=&\lambda_{x},\quad \text{i is even.} \nonumber
\end{eqnarray}
Since the chains on different lines $i$ are not coupled, $[\hat{K}_m, \hat{K}_n]=0$ and each Ising chain can be independently exactly solved by means of usual techniques involving Jordan-Wigner transformation, a Fourier transform and finally a Bogoliubov transformation\cite{Lieb1961}. Moreover, the ground state of Eq.(\ref{isingH}) is the tensor product of the ground states for each Ising chain on the line $i$, that is,
\begin{eqnarray}
|\Psi\rangle=\otimes_{i=1}^{2N}|\Psi_{i}\rangle
\label{factorized}
\end{eqnarray}
in which $|\Psi_{i}\rangle$is the ground state of the $i$-th Ising chain.
From now, we call the representation in terms of the effective spin operators $\tau^z_s, \tau^x_s$ the `$\tau$-picture'. The $\tau$ spins live on the sites of the lattice and the dual lattice, while the `$\sigma$-picture' refers to the Hamiltonian Eq.(\ref{Hsigma}) written in terms of the original spins $\sigma$ living on the bonds of the lattice, see Fig.\ref{lattice2}.
What we need to pay attention to is that there is one constraint in each Ising chain caused by periodic boundary condition, that is $\tau^{x}_{s^{i}_{N}}\tau^{x}_{s^{i}_1}=\sigma^z_{<0,1>^i}$ for any row index i.
It corresponds to the constraint in '$\sigma$-picture':
\begin{eqnarray}
\prod_{j=1}^{N}&\sigma^{z}_{<j-1,j>^{2k-1}}&=1,\nonumber\\
\prod_{j=1}^{N}&\sigma^{x}_{<j-1,j>^{2k}}&=1,\quad k=1,2,...,N.
\end{eqnarray}
Notice that the operators $w^{z}_{k}=\prod_{j=1}^{N}\sigma^{z}_{<j-1,j>^{2k-1}}$ and $w^{x}_{k}=\prod_{j=1}^{N}\sigma^{x}_{<j-1,j>^{2k}}$ are the topological operators in TCM, which means $W^z_2$ and $W^x_1$. From now on whenever the operator $W^x_1$($W^z_2$) occurs we mean that it is just an arbitrary $w^{x}_{k}$($w^{z}_{k}$).
 They all commute with $H$, so we have $2N$ conserved quantities. If we denote the whole Hilbert space as $\mathscr{H}$ ('$\sigma$-picture') which dimension is $2^{2N^{2}}$, we can choose the sector
\begin{eqnarray}
\mathscr{H^{\prime}}=\{&&|\Psi\rangle\in\mathscr{H}\mid w^{z}_{k}|\Psi\rangle=|\Psi\rangle, ~w^{x}_{k}|\Psi\rangle=|\Psi\rangle,\nonumber\\
~&&k=1,2,...,N \},
\end{eqnarray}
whose dimension is $2^{2N^2-2N}$. In this sector, the product $\prod_{j=1}^{N}A_{s^{2k-1}_j}$ and $\prod_{j=1}^{N}B_{s^{2k}_j}$ also equal to identity. We can write these constraints in the '$\tau$-picture' as
\begin{eqnarray}
\prod_{j=1}^{N}&\tau^{z}_{s^{2k-1}_j}&=1,\nonumber\\
\prod_{j=1}^{N}&\tau^{z}_{s^{2k}_j}&=1,\quad k=1,2,...,N.
\end{eqnarray}
These constraints together with periodic boundary condition give the corresponding sector $\mathscr{H^{\prime}}$ in '$\tau$-picture'.

\subsection{Derivation of formula for the purity}
In this section, we find a general formula to compute the purity for a generic state in $\mathscr{H^{\prime}}$ that is factorizable in the product on different lines $i$ as in Eq.(\ref{factorized}). In this way, this formula can be used to compute the purity of every eigenstate of the Hamiltonian Eq.(\ref{Hsigma}) or of the time evolution, including the one induced by a sudden quantum quench.

First, we need to choose a reference state which is a vector in $\mathscr{L}$ and also in $\mathscr{H^{\prime}}$. As it is immediate to verify,  the following state is a vector in $\mathscr{L}$:
\begin{eqnarray}
\label{eq:2}
|0^\prime\rangle&\equiv&|G|^{-1/2}\sum_{g\in G}g|\Uparrow\rangle
\end{eqnarray}
where $|\Uparrow\rangle$ is the state with all spins pointing up in the $z$-basis, namely $\sigma^{z}_{i}|\Uparrow\rangle=|\Uparrow\rangle,$ $\forall i$. $G$ is the group generated by the $N^2-1$ independent star operators $A_{s}$.
The state Eq.(\ref{eq:2}) is  the simultaneous eigenstate of $W^{z}_{1}$ and $W^{z}_{2}$ with eigenvalue $1$. However, it is not the the vector in the sector $\mathscr{H^{\prime}}$. It is more convenient to  choose, as reference state,  the following state in $\mathscr{H^{\prime}}$ (also in $\mathscr{L}$ ):
\begin{eqnarray}
\label{eq:3}
|0\rangle=\frac{1+W^{x}_{1}}{\sqrt{2}}|0^\prime\rangle=(2|G|)^{-1/2}\sum_{g\in G} g(1+W^{x}_{1})|\Uparrow\rangle.\nonumber\\
\end{eqnarray}
This state belongs to a different topological sector, being the eigenstate of $W^{x}_{1}$ and $W^{z}_{2}$ with eigenvalue $1$. Any state in $\mathscr{H^{\prime}}$ can be written as:
\begin{eqnarray}
\label{eq:4}
|\Psi\rangle=\sum_{x\in X}\sum_{z\in Z}b(xz)zx|0\rangle.
\end{eqnarray}
The group $X(Z)$ has the tensor product form $X=\otimes_{k=1}^{N}X_{k}$ $(Z=\otimes_{k=1}^{N}Z_{k})$. One defines strings of $x-$type as the strings running on the dual lattice and connecting the centres of plaquettes, and acting as $\hat{\sigma}^x$ on all the spins intersected by the string. Likewise, the strings of type $z$ act like $\hat{\sigma}^z$ on all the spins traversed by strings running on the links of the lattice and connecting the sites of the lattice, see Fig.\ref{lattice}. The elements of each $X_{k}$($Z_{k}$) are the open strings of $x(z)$-type, mod $\{1,W^{x}_{1}\}$ ($\{1,W^{z}_{2}\}$), lying on the $2k$th\;$\left((2k-1)th\right)$ row. The number of open strings' endpoints is even, so the number of independent open strings is $(\sum_{m=1}^{[N/2]}\left(\begin{subarray}{c}N \\2m\end{subarray}\right))^{2N}=2^{2N^2-2N}$, which is conform to the dimension of $\mathscr{H^\prime}$.

As we noticed above, in the view of `$\tau$-picture', $|\Psi\rangle$ and $|0\rangle$ have tensor product form: $|\Psi_\tau\rangle=|\Psi_1\rangle\ldots|\Psi_{2N^2}\rangle$ and
$|0_\tau\rangle=|0_1\rangle\ldots|0_{2N^{2}}\rangle$.
Let us  introduce the following notation for $|\psi\rangle$ and $|\phi\rangle$ as follows:
\begin{eqnarray}\label{phiandpsi}
|\Psi_\tau\rangle:\left\{
\begin{aligned}
|\psi\rangle&=&&|\Psi_1\rangle|\Psi_3\rangle\ldots|\Psi_{2N^2-1}\rangle\\
|\phi\rangle&=&&|\Psi_2\rangle|\Psi_4\rangle\ldots|\Psi_{2N^2}\rangle
\end{aligned}
\right.
\end{eqnarray}
and
\begin{eqnarray}
|0_\tau\rangle:\left\{
\begin{aligned}
|0_\psi\rangle&=&&|0_1\rangle|0_3\rangle\ldots|0_{2N^2-1}\rangle\\
|0_\phi\rangle&=&&|0_2\rangle|0_4\rangle\ldots|0_{2N^2}\rangle.
\end{aligned}
\right.
\end{eqnarray}
The following otrhonormality conditions are easily proven in the `$\tau$-picture':
\begin{eqnarray}
\label{oxz}
\langle0|xz|0\rangle&=& \langle0_{\psi}|\tau(x)|0_{\psi}\rangle\langle0_{\phi}|\tau(z)|0_{\phi}\rangle\nonumber\\
 &=& \delta_{z,\mathbbm{1}_{Z}}\delta_{x,\mathbbm{1}_{X}},~\text{$\forall x\in X,\forall z\in Z$},
\end{eqnarray}
and
\begin{eqnarray}
\label{bxz}
b(xz)=\langle0|xz|\Psi\rangle=\langle0_{\psi}|\tau(x)|\psi\rangle\langle0_{\phi}|\tau(z)|\phi\rangle\equiv b(x)b(z)\nonumber\\
\end{eqnarray}
where $\tau$(x)($\tau$(z)) is the operator mapping from `$\sigma$-picture' to the `$\tau$-picture'.
Combining Eqs.(\ref{eq:2}), (\ref{eq:3}) and (\ref{eq:4}),we  get:
\begin{eqnarray}
\label{psi1}
|\Psi\rangle=(2|G|)^{-1/2}\sum_{x\in X}\sum_{g\in G}\sum_{z\in Z}b(xz)zxg(1+W_1^x)|\Uparrow\rangle.\nonumber\\
\end{eqnarray}
Now we introduce a new group $Y$ for convenience of the later derivation which is defined as
\begin{eqnarray}
\label{Y}
Y=X\times G\times \{1,W_1^x\}.
\end{eqnarray}
The generators of $Y$ are of course all the generators of $X$, $G$ and $W_1^x$, but we can also give a different description that will come useful later. As one can easily verify, the group can be generated by two types of operators: (1) all the $\sigma^x$ operators lying on the vertical lines ( even rows ), and (2) all the open strings formed by $\sigma^x$ operators lying on the horizon lines ( odd rows ). For example, the generators in (2k-1)th row is the operator $\sigma^{x}_{<i,i+1>^{2k-1}}\sigma^{x}_{<i+1,i+2>^{2k-1}}$, $i=1,2,\ldots,N-1$. The numbers of generators belonging to the first  type is $N^2$ while for the type two is $N(N-1)$. Thus the order of $Y$ is $2^{2N^2-N}$ which is of course identical to $|X|\times|G|\times|\{1,W_1^x\}|=2^{2N^2-N}$. So for any $x\in X$ and $g\in G\times\{1,W_1^x\}$ we have following relationships
\begin{eqnarray}\label{Y2}
\forall x,g \quad \exists y\in Y,\quad s.t.\quad y=xg; \nonumber\\
b(xz)=b(gxz)=b(yz).
\end{eqnarray}
Where we used Eq.($\ref{bxz}$) and the fact that $g|0\rangle=|0\rangle$ in the second equation.
By combining the Eqs.(\ref{Y})(\ref{Y2}) we rewrite  Eq.(\ref{psi1}) as
\begin{eqnarray}
\label{psi2}
|\Psi\rangle=(2|G|)^{-1/2}\sum_{z\in Z}\sum_{y\in Y}b(yz)zy|\Uparrow\rangle.
\end{eqnarray}
Note that whether the operators $z$ and $y$ commute or not depending on the common links they shared. If they share even (odd) links, they commute( don't commute). The parity of the shared links number is denoted as $z\cap y$, namely:
\begin{eqnarray}
zy=yz(-1)^{z\cap y}.
\end{eqnarray}
Together with the fact that $z|\Uparrow\rangle$ Eq.$(\ref{psi2})$ turn out to be:
\begin{eqnarray}
|\Psi\rangle=(2|G|)^{-1/2}\sum_{z\in Z}\sum_{y\in Y}(-1)^{y\cap z}b(yz)y|\Uparrow\rangle.
\end{eqnarray}
Now, let us write down the corresponding density operator. It reads:
\begin{eqnarray}
\rho=&&(2|G|)^{-1}\sum_{z,z^\prime \in Z}\sum_{y,y^\prime \in Y} \bar{b}(z^\prime y^\prime)b(yz)\nonumber\\
&&\times(-1)^{y\cap z+y^\prime\cap z^\prime}|y_Ay_B\rangle\langle y^\prime_Ay^\prime_B|
\end{eqnarray}
where we have adopted the notation: $y|\Uparrow\rangle=y_A|\Uparrow_A\rangle\otimes y_B|\Uparrow_B\rangle=|y_Ay_B\rangle$.
The reduced density operator to subsystem $A$ is obtained by tracing over the $B$ part
\begin{eqnarray}
\label{rhoA1}
\rho_A=&&(2|G|)^{-1}\sum_{z,z^\prime \in Z}\sum_{y, y\prime\in Y}\bar{b}(z^\prime y^\prime )b(yz)\nonumber\\
&&\times(-1)^{y\cap z+y^\prime\cap z^{\prime}}|y_A\rangle\langle y^\prime_A|\langle y_B y^\prime_B\rangle.
\end{eqnarray}
Since we are summing over all the elements of the group, we can relabel the elements in the sum as $ y^\prime=y\tilde{y}$ , to rewrite Eq.($\ref{rhoA1}$) as
\begin{eqnarray}
\rho_A=&&(2|G|)^{-1}\sum_{z,z^\prime \in Z}\sum_{y,\tilde{y}\in Y}\bar{b}(z^\prime y\tilde{y})b(yz)\nonumber\\
&&\times(-1)^{y\cap z+y\tilde{y}\cap z^{\prime}}|y_A\rangle\langle y_A\tilde{y}_A|\langle\tilde{y}_B\rangle.
\end{eqnarray}
Note that $\langle\tilde{y}_B\rangle$ is nonzero only when $\tilde{y}_B=\mathbbm{1}_{B}$. We now introduce the subgroups $Y_A\in Y$ and $Y_B\in Y$, \begin{eqnarray}
\label{YAB}
Y_A&\equiv&\{y\in Y|y=y_A\otimes\mathbbm{1}_{B}\},\\
Y_B&\equiv&\{y\in Y|y=\mathbbm{1}_{A}\otimes y_B\}.
\end{eqnarray}
Finally we get the reduced operator in form of
\begin{eqnarray}
\rho_A=&&(2|G|)^{-1}\sum_{z,z^\prime \in Z}\sum_{\substack{y\in Y\\ \tilde{y}\in Y_A}}\bar{b}(z^\prime y \tilde{y})b(yz)\nonumber\\
&&\times(-1)^{y\cap z+y\tilde{y}\cap z^{\prime}}|y_A\rangle\langle y_A\tilde{y}_A|.
\end{eqnarray}

{Let us now make a remark about the topological sector used in this derivation. The state $|\Psi\rangle$ we are interested in is a  state away from equilibrium after quantum quench, that is,  $|\Psi\rangle =e^{-iH(\lambda)t}|\Psi(0)\rangle$. The initial state $|\Psi(0)\rangle$ is a ground state of toric code Hamiltonian $H(\lambda=0)$ which we prepared at t=0. In the derivation, the state  $|\Psi(0)\rangle$ ( also $|\Psi\rangle$ ) is constrained to the sector $\mathscr{H^{\prime}}$, which is the eigenspace of $W^x_1=1$ and $W^z_2=1$, that is $|\Psi(0)\rangle=|0\rangle$. However, topological entropy is not affected by this restriction. Following \cite{hamma:2005b}, we can show that the reduced density matrix $\rho_A=\mbox{Tr}_B[|\Psi\rangle\langle\Psi|]$ is independent on the topological sector, and thus there is no loss of generality in fixing it. Indeed,  by denoting the following 4 states $|\xi_{ij}\rangle, i,j=0,1$ as a basis in the ground state manifold $\mathscr{L}$:
\begin{eqnarray}
|\xi_{ij}\rangle=(W^z_1)^i(W^x_2)^j|0\rangle ,
\end{eqnarray}
we see that they  satisfy $W^x_1|\xi_{ij}\rangle=(-1)^i|\xi_{ij}\rangle$ and $W^z_2|\xi_{ij}\rangle=(-1)^j|\xi_{ij}\rangle$. An arbitrary state in $\mathscr{L}$ can be written as
\begin{eqnarray}
|\tilde\xi\rangle=\sum_{i,j=0}^1 \alpha_{ij}|\xi_{ij}\rangle.
\end{eqnarray}
where $\sum_{i,j=0}^1 |\alpha_{ij}|=1$.
After the same procedure showed in eq.(\ref{eq:4}), we can get the corresponding $|\tilde\Psi_{ij}\rangle$  and also $|\tilde\Psi\rangle$ as
\begin{eqnarray}
|\Psi_{ij}\rangle=\sum_{x\in X}\sum_{z\in Z}b(xz)zx|\xi_{ij}\rangle.
\end{eqnarray}
and
\begin{eqnarray}
|\tilde\Psi\rangle=\sum_{i,j=0}^1 \alpha_{ij}|\Psi_{ij}\rangle
\end{eqnarray}
where $W^x_1|\Psi_{ij}\rangle=(-1)^i|\Psi_{ij}\rangle$ and $W^z_2|\Psi_{ij}\rangle=(-1)^j|\Psi_{ij}\rangle$. The reduced density matrix of $|\tilde\Psi\rangle$ is
\begin{eqnarray}
\tilde\rho_A=\sum_{i,j,k,l=0}^1\alpha_{ij}\alpha_{kl}^\ast \mbox{Tr}_B[|\Psi_{ij}\rangle\langle\Psi_{kl}|].
\end{eqnarray}
We can thus prove that $\mbox{Tr}_B[|\Psi_{ij}\rangle\langle\Psi_{kl}|]=\delta_{ij,kl}\mbox{Tr}_B[|\Psi_{00}\rangle\langle\Psi_{00}|]=\rho_A$. A similar proof was showed in \cite{hamma:2005b},  where the fact that contractible loops can not generate non-contractible loop was used. Noticing that contractible loops and open strings also can not generate non-contractible loop, the proof can be directly generalized.
We therefore have
\begin{eqnarray}
\tilde\rho_A=\rho_A.
\end{eqnarray}
$|\tilde\Psi\rangle$ belongs to the space $\mathscr{\tilde H}$, which is defined as
\begin{eqnarray}
\mathscr{\tilde H}=\{&&|\Psi\rangle\in\mathscr{H}\mid \prod_{j=1}^{N}A_{s^{2k-1}_j}|\Psi\rangle=|\Psi\rangle,\nonumber\\
&&\prod_{j=1}^{N}B_{s^{2k}_j}|\Psi\rangle=|\Psi\rangle,~k=1,2,...,N \}.
\end{eqnarray}
Paying attention to the global constraint of $\prod_sA_s=1$ and $\prod_pB_p=1$, we have dim($\mathscr{\tilde H}$)=4dim($\mathscr{H^\prime}$). Also we have $\mathscr{H^\prime}\subset\mathscr{\tilde H}$ and $\mathscr{L}\subset\mathscr{\tilde H}$ ( note that $\mathscr{L}\nsubset\mathscr{H^\prime}$).}

\begin{widetext}
Now we move on to the calculation of the purity of $\rho_ A$, which is $P=\mbox{Tr}[\rho_A^2]$, follows directly as
\begin{eqnarray}
P=&&(2|G|)^{-2}\sum_{\substack{z_1,z_2\in Z\\z_1^\prime,z_2^\prime\in Z}}\sum_{\substack{y_1,y_2\in Y\\ \tilde{y}_1,\tilde{y}_2\in Y_A}}\bar{b}(z_1^\prime y_1\tilde{y}_1)b(y_1z_1)\bar{b}(z_2^\prime y_2\tilde{y}_2)b(y_2z_2)\nonumber\\
&&\times(-1)^{y_1\cap z_1+y_1\tilde{y}_1\cap z_1^\prime+y_2\cap z_2+y_2\tilde{y}_2\cap z_2^\prime}\langle y_{1A}\tilde{y}_{1A}y_{2A}\rangle\langle y_{2A}\tilde{y}_{2A}y_{1A}\rangle.
\end{eqnarray}
Note that the term $\langle y_{1A}\tilde{y}_{1A}y_{2A}\rangle\langle y_{2A}\tilde{y}_{2A}y_{1A}\rangle$ imposes  two constrains: (1) $\tilde{y}_1=\tilde{y}_2$; (2) $y_2=y_1\tilde{y_1}\bar{y}$ where $\bar{y}\in Y_B$. Thus the purity formula can be simplified as
\begin{eqnarray}
\label{P1}
P=&&(2|G|)^{-2}\sum_{\substack{z_1,z_2\in Z\\z_1^\prime,z_2^\prime\in Z}}\sum_{\substack{y\in Y\\\tilde{y}\in Y_A\\ \bar{y}\in Y_B}}\bar{b}(z_1^\prime y\tilde{y})b(yz_1)\bar{b}(z_2^\prime y\bar{y})b(y\tilde{y}\bar{y}z_2)\nonumber\\
&&\times (-1)^{y\cap z_1+ y\tilde{y}\cap z_1^\prime +y\tilde{y}\bar{y}\cap z_2 +y\bar{y}\cap z_2^\prime}.
\end{eqnarray}
For further simplification, we rewrite the last term as
\begin{eqnarray}
\label{phase}
(-1)^{y\cap z_1+ y\tilde{y}\cap z_1^\prime +y\tilde{y}\bar{y}\cap z_2 +y\bar{y}\cap z_2^\prime}=(-1)^{\tilde{y}\cap z_1^\prime +\tilde{y}\cap z_2}(-1)^{y\cap z_1z_1^\prime z_2z_2^\prime}(-1)^{\bar{y}\cap z_2z_2^\prime}.
\end{eqnarray}
 The above equality can be easily proven by the fact that:
\begin{eqnarray}
y_1\ldots y_k z_1\ldots z_l&=&z_1\ldots z_l y_1\ldots y_k(-1)^{g_1\ldots g_k\cap z_1\ldots z_l};\nonumber\\
y_1\ldots y_k z_1\ldots z_l&=&z_1\ldots z_l y_1\ldots y_k\prod_{i=1}^k\prod_{j=1}^l (-1)^{y_i\cap z_j}.
\end{eqnarray}
The first equation is deduced as we commute the $(y_1\ldots y_k)$ and $(z_1\ldots z_l)$ as two operators while the second equation we commute each $y_i$ and $z_j$ at a time.
Now recall that   $\bar{b}(z_1^\prime y\tilde{y})b(y\tilde{y}\bar{y}z_2)$ is equal to $\langle\Psi|z_1^\prime y\tilde{y}|0\rangle\langle0|y\tilde{y}\bar{y}z_2|\Psi\rangle$, we have
\begin{eqnarray}
\bar{b}(z_1^\prime y\tilde{y})b(y\tilde{y}\bar{y}z_2)(-1)^{\tilde{y}\cap z_1^\prime +\tilde{y}\cap z_2}&=&\langle\Psi|z_1^\prime y\tilde{y}|0\rangle\langle0|y\tilde{y}\bar{y}z_2|\Psi\rangle(-1)^{\tilde{y}\cap z_1^\prime +\tilde{y}\cap z_2}\nonumber \\
&=&\langle\Psi|\tilde{y}z_1^\prime y|0\rangle\langle0|y\bar{y}z_2\tilde{y}|\Psi\rangle.
\end{eqnarray}
Where we have employed that every two elements in group Y commute. Combining this equation, Eq.$(\ref{P1})$ is simplified to be
\begin{eqnarray}
\label{P2}
P=&&(2|G|)^{-2}\sum_{\substack{z_1,z_2\in Z\\z_1^\prime,z_2^\prime\in Z}}\sum_{\substack{y\in Y\\ \tilde{y}\in Y_A\\ \bar{y}\in Y_B}}\langle\Psi|\tilde{y}z_1^\prime  y|0\rangle\langle 0|yz_1|\Psi\rangle\langle\Psi|z_2^\prime y\bar{y}|0\rangle\langle 0|y\bar{y}z_2\tilde{y}|\Psi\rangle\\ \nonumber
&&\times(-1)^{y\cap z_1z_1^\prime z_2z_2^\prime (-1)\bar{y}\cap z_2z_2^\prime}.
\end{eqnarray}
\end{widetext}
Notice that the above formula is written in the `$\sigma$-picture'. In the following, we will obtain the exact state $|\Psi\rangle$ in the $\tau$ picture. Therefore, in order to proceed to  further calculations, we  first need to map this formula in `$\sigma$-picture' to `$\tau$-picture'. We know how to map the group $Z$ to `$\tau$-picture' as we discussed earlier,  but how about the group $Y$?
One has to remember that $Y$ is generated by two types of operators. Again, the first type is all the $\sigma^x$ operators lying on the vertical lines ( even rows ). These operators form a group  $X^\prime$, which is homomorphic to the group $X$ . Notice  that both $X$ and $X^\prime$ possess tensor product form of each even rows and the homomorphic mapping from $X^\prime$ to $X$ in each rows is $2$ to $1$ (since $X$ is a group containing only open strings  while $X^\prime$ containing open strings and a non-contractible  closed string in each row. In $\mathscr{H^{\prime}}$, this non-contractible closed string  acts as identity operator){, the homomorphic mapping from  $X^\prime$ to X is $2^N$ to 1 for the number of even rows is N. So the order of $X^\prime$ is $|X^\prime|=2^{N^2}$ while $|X|=2^{N^2-N}$. Clearly $X^\prime$ can be mapped to the '$\tau$-picture'.
The second  type is all the open strings of $\sigma^x$ operators lying on the horizontal lines ( odd rows ). In order to map them to the '$\tau$-picture', we exploit again the relabelling in the sum over all the elements of a group and  replace them by the operators forming the group $G^\prime=Y/X^\prime$. This is a group formed by some contractible loops in dual lattice of $\sigma^x$ type (and of course it is a subgroup of G). Precisely, $G^\prime=G\times\{1,W^x_1\}/\prod_{k=1}^N \{1,w^x_k\}$. The description in '$\tau$-picture' is more clear: $G^\prime$ has the tensor product form $G^\prime=\otimes_{k=1}^N G^\prime_k$ and each $G^\prime_k$ is generated by $2^{N-1}$ independent $\tau^z_{s^{2k-1}_j}$ with constrain $\prod_{j=1}^{N}\tau^z_{s^{2k-1}_j}=1$  in the (2k-1)th row. Or we can say that $G^\prime$ is generated by the open strings of $\tau^z$ operators lying on the odd rows. So $|G^\prime|=2^{N^2-N}$ and $|Y|=|G^\prime||X^\prime|=2^{2N^2-N}$ coinciding with the former discussion. Finally the group $Y$ can be written as
\begin{eqnarray}
Y=G^\prime\times X^\prime.
\end{eqnarray}

The next step is to rewrite $Y_A$ and $Y_B$. This part is a little difficult because of the constraints on the boundary of subsystems $A$ and $B$, which is showed in  Figs.\ref{partialGAGB}. One can verify that the relationships showed in Table \ref{groupY} hold.
\begin{table}[t]
\begin{tabular}{|l|l|l|}
\hline
$\forall y\in Y$ & $\exists$ $g\in G^\prime$, $x\in X^\prime$ & s.t. $y=gx$\\
\hline
$\forall \tilde{y}\in Y_A$ & $\exists$ $\tilde{g}\in G^\prime_A$, $\partial\tilde{g}\in\partial G^\prime_A$, $\tilde{x}\in X^\prime_A$ & s.t. $\tilde{y}=\tilde{g}\tilde{x}\partial\tilde{g}\partial\tilde{x}(\partial\tilde{g})$\\
\hline
$\forall \bar{y}\in Y_B$ & $\exists$ $\bar{g}\in G^\prime_B$, $\partial\bar{g}\in\partial G^\prime_B$, $\bar{x}\in X^\prime_B$ & s.t. $\bar{y}=\bar{g}\bar{x}\partial\bar{g}\partial\bar{x}(\partial\bar{g})$\\
\hline
\end{tabular}
\caption{Decompositions of Group $Y$, $Y_A$ and $Y_B$.}\label{groupY}
\end{table}
In this table, the groups $G^\prime$, $G^\prime_A$, $G^\prime_B$, $\partial G^\prime_A$ and $\partial G^\prime_B$ are all subgroups of $G$. $\tilde{x}(\partial\tilde{g})$ and $\partial\bar{x}(\partial\bar{g})$ are the functions of $\partial\tilde{g}$ and $\partial\bar{g}$ respectively.
Moreover,
$G^\prime_A\subset G^\prime$ is generated by all the independent star operators that act solely on subsystem $A$, while
$G^\prime_B\subset G^\prime$ is generated by all the independent star operators that act solely on subsystem $B$.
The generators of $\partial G^\prime_A$ and $\partial G^\prime_B$ are showed in Fig.\ref{partialGAGB}. They depend upon the shape of the subsystem A and we choose subsystem (2) to illustrate and you can get them for subsystem (1), (3) and (4).
\begin{figure*}[t]
\begin{center}
\includegraphics[width=1\textwidth]{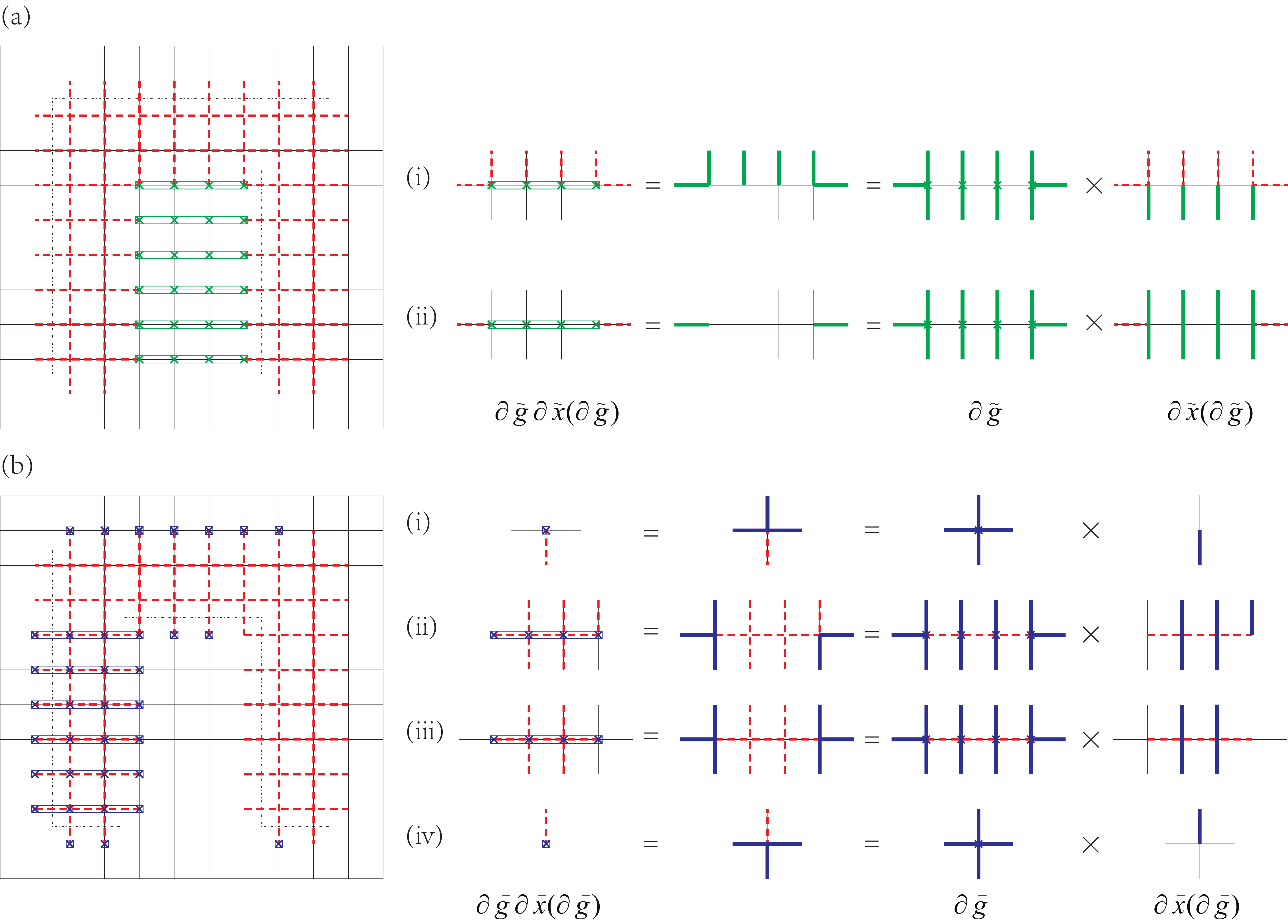}\\
\caption{ Illustration of (a) $\partial \tilde{g}\partial \tilde{x}(\partial \tilde{g})$ operators that $\partial \tilde{g}\in \partial G^\prime_A$, (b) $\partial\bar{g}\partial\bar{x}(\partial\bar{g})$ operators that $\partial\bar{g}\in\partial G^\prime_B$ in subsystem (2) with R=8, r=2. These collective operators are denoted as the (a) green crosses with rectangle block, (b) blue crosses with square or rectangle block. The (a) green (b) blue solid lines on the right side are the $\sigma^x$ operators on the edges. Each generator of the $\partial G^\prime_A$ is the product of the star operators on the cross marked sites in each rectangle block. (a) $\partial\tilde{x}$ is the function of $\partial\tilde{g}$ so that $\partial \tilde{g}\partial \tilde{x}(\partial \tilde{g})$ acts solely on A. There are 2 types of $\partial \tilde{g}\partial \tilde{x}(\partial \tilde{g})$ operators in subsystem (2). (b)  The generators of $\partial G^\prime_B$ are: (i),(iv) the star operator on the cross marked site in each of the square blocks; (ii),(iii) the product of star operators on the cross marked sites in each of rectangle blocks. $\partial\tilde{x}$ is the function of $\partial\bar{g}$ so that $\partial\bar{g}\partial\bar{x}(\partial\bar{g})$ acts solely on B. There are 4 type of $\partial \bar{g}\partial \bar{x}(\partial \bar{g})$ operators in the subsystem (2). }\label{partialGAGB}
\end{center}
\end{figure*}
\begin{widetext}
Now we can  map the spin operators in formula of purity Eq.($\ref{P2}$) to the `$\tau$-picture'. From now on, we will not   distinguish the notations of all the groups concerned in the two pictures. The purity in the $\tau$ picture finally reads:
\begin{eqnarray}
P=\sum_{\substack{\partial\tilde{g}\in\partial G^\prime_A\\ \partial\bar{g}\in\partial G^\prime_B}}P_1(\partial\tilde{g},\partial\bar{g})P_2(\partial\tilde{g},\partial\bar{g})
\end{eqnarray}
where, recalling $|\psi\rangle$ and $|\phi\rangle$  defined in Eq.($\ref{phiandpsi}$), we have
\begin{eqnarray}
\label{P_1}
P_1(\partial\tilde{g},\partial\bar{g})=&&\sum_{\substack{z_1,z_2\in Z\\ z_1^\prime,z_2^\prime\in Z}}\sum_{\substack{g\in G^\prime\\ \tilde{g}\in G_A^\prime\\ \bar{g}\in G_B^\prime}}(2|G|)^{-2}\langle\psi|\tilde{g}\partial\tilde{g}z_1^\prime g|0_\psi\rangle\langle 0_\psi|gz_1|\psi\rangle\langle\psi|z_2^\prime g\bar{g}\partial\bar{g}|0_\psi\rangle\langle 0_\psi|g\bar{g}\partial\bar{g}z_2\tilde{g}\partial\tilde{g}|\psi\rangle\\ \nonumber
&&\times(-1)^{g\cap z_1z_1^\prime z_2 z_2^\prime}(-1)^{\bar{g}\cap z_2 z_2^\prime}(-1)^{\partial\bar{g}\cap z_2 z_2^\prime}
\end{eqnarray}
and
\begin{eqnarray}
\label{P_2}
P_2(\partial\tilde{g},\partial\bar{g})=\sum_{\substack{x\in X^\prime\\ \tilde{x}\in X_A^\prime\\ \bar{x}\in X_B^\prime}}\langle \phi|\tilde{x}\partial\tilde{x}(\partial\tilde{g})x|0_\phi\rangle\langle 0_\phi|x|\phi\rangle\langle\phi|x\bar{x}\partial\bar{x}(\partial\bar{g})|0_\phi\rangle\langle 0_\phi|x\bar{x}\partial\bar{x}(\partial\bar{g})\tilde{x}\partial\tilde{x}(\partial\tilde{g})|\phi\rangle.
\end{eqnarray}

There are two point to notice. (i) The phase term don't appear in $P_2$ because of the fact that $x$ and $z$ live in even and odd rows respectively so they always commute. (ii) The notation  $g\cap z$ in $P_1$ means the parity of the number of  common sites   shared by $g$ and $z$ in '$\tau$-picture'.
 Remembering that $g\cap z$ is the parity of common links in '$\sigma$-picture' as we have introduced before.
The above expression can be simplified. Let us start with
$P_1$. First, notice  the fact that for any $g\in G$, we have $g|0_\psi\rangle=|0_\psi\rangle$, so the $g$ type of operators in Eq.($\ref{P_1}$) are absorbed by the $|0_\psi\rangle$:
\begin{eqnarray}
\label{P1term1}
&&\langle\psi|\tilde{g}\partial\tilde{g}z_1^\prime g|0_\psi\rangle\langle 0_\psi|gz_1|\psi\rangle\langle\psi|z_2^\prime g\bar{g}\partial\bar{g}|0_\psi\rangle\langle 0_\psi|g\bar{g}\partial\bar{g}z_2\tilde{g}\partial\tilde{g}|\psi\rangle\nonumber\\
=&&\langle\psi|\tilde{g}\partial\tilde{g}z_1^\prime|0_\psi\rangle\langle 0_\psi|z_1|\psi\rangle\langle\psi|z_2^\prime |0_\psi\rangle\langle 0_\psi|z_2\tilde{g}\partial\tilde{g}|\psi\rangle.
\end{eqnarray}
Next, we work on the last phase term in Eq.($\ref{P_1}$). We can prove the following equation:
\begin{eqnarray}
\label{phase2}
\sum_{g\in G^\prime_R}(-1)^{g\cap z}=\left\{
\begin{aligned}
&|G^\prime_R| &\quad  z\in Z_{\bar{R}}\\
&0   &\quad   z\notin Z_{\bar{R}}
\end{aligned}
\right.
\end{eqnarray}
where $Z_{\bar{R}}$ is defined as: $Z_{\bar{R}}\equiv\{z\in Z|\forall g\in G_R , zg=gz\}$. The proof goes as follows.
If $\exists a\in G^\prime_R$ s.t. $az=-za$. Define quotient group $G_a\equiv G^\prime_{R}/\{1,a\}$,  thus $G^\prime_R=\{ G_a, aG_a\}$.
Then $\sum_{g\in G^\prime_R}(-1)^{g\cap z}=\sum_{g\in G_a}(-1)^{g\cap z}+\sum_{g\in aG_a}(-1)^{g\cap z}$. The second term equal to $\sum_{g\in G_a}(-1)^{ag\cap z}=\sum_{g\in G_a}(-1)^{g\cap z}(-1)^{a\cap z}=-\sum_{g\in G_a}(-1)^{g\cap z}$, since $az=-za$. So $\sum_{g\in G^\prime_R}(-1)^{g\cap z}=0$. If $\forall g\in G^\prime_R$ satisfies $gz=zg$, $(-1)^{g\cap z}=1$, so $\sum_{g\in G^\prime_R}(-1)^{g\cap z}=\sum_{g\in G^\prime_R}1=|G^\prime_R|$.
Combining Eqs.($\ref{P1term1}, \ref{phase2}$) in Eq.($\ref{P_1}$) we get the relations $z_1 z_1^\prime z_2 z_2^\prime=1$ and $z_2 z_2^\prime=\tilde{z}\in Z_{\bar{B}}$. We prefer to rename the group $Z_{\bar{B}}$ as $Z_A$, thus Eq.($\ref{P_1}$) is simplified as
\begin{eqnarray}
P_1=&&\frac{|G^\prime||G_B^\prime|}{(2|G|)^2}\sum_{\substack{z_1,z_2\in Z\\ \tilde{z}\in Z_A}}\sum_{\tilde{g}\in G_{A}^\prime}\langle\psi|\tilde{g}\partial\tilde{g}\tilde{z}z_1|0_\psi\rangle\langle 0_\psi|z_1|\psi\rangle\nonumber\\
&&\times\langle\psi|\tilde{z}z_2|0_\psi\rangle\langle 0_\psi|z_2\tilde{g}\partial\tilde{g}|\psi\rangle(-1)^{\partial\bar{g}\cap\tilde{z}}.
\end{eqnarray}
Noticing the fact that $\sum_{z\in Z}z|0_\psi\rangle\langle 0_\psi|z=1$ and $2|G|=2^N |G^\prime|$, we finally get:
\begin{eqnarray}
\label{P1f}
P_1(\partial\tilde{g},\partial\bar{g})=\frac{|G_B^\prime|}{2^{2N}|G^\prime|}\sum_{\tilde{z}\in Z_A}\sum_{\tilde{g}\in G^\prime_A}|\langle\psi|\tilde{g}\partial\tilde{g}\tilde{z}|\psi\rangle|^2(-1)^{\partial\bar{g}\cap\tilde{z}}.\nonumber\\
\end{eqnarray}

Now, let us take care of $P_2$. Just like the group $G^\prime$ defined in odd rows, we define the corresponding group in even rows as $H^\prime$. In '$\tau$-picture', $H^\prime$ has the tensor product form $H^\prime=\otimes_{k=1}^N H^\prime_k$ and each $H^\prime_k$ is generated by $2^{N-1}$ independent $\tau^z_{s^{2k}_j}$ with constrain $\prod_{j=1}^{N}\tau^z_{s^{2k}_j}=1$  in the (2k)th row. Now we rewrite the $|0_\phi\rangle$ as
\begin{eqnarray}
|0_\phi\rangle=|H^\prime|^{-1/2}\sum_{h\in H^\prime}h|\tilde{0}\rangle
\end{eqnarray}
where $|\tilde{0}\rangle$ has the tensor product form $|\tilde{0}\rangle\equiv|\tilde{0}_\phi\rangle=\otimes_{k=1}^{N}|\tilde{0}_{2k}\rangle$. It satisfies  $x|\tilde{0}\rangle=|\tilde{0}\rangle$ for all $x\in X$ (it holds for $x\in X^\prime$) and $\prod_{j=1}^{N}\tau^z_{s^{2k}_j}|\tilde{0}_{2k}\rangle=|\tilde{0}_{2k}\rangle$ for any k.
Substitute the equation into Eq.($\ref{P_2}$):
\begin{eqnarray}
P_2=|H^\prime|^{-2}\sum_{\substack{h_1,h_2\in H^\prime\\ h_1^\prime,h_1^\prime\in H^\prime}}\sum_{\substack{x\in X^\prime\\ \tilde{x}\in X_A^\prime\\ \bar{x}\in X_B^\prime}}\langle\phi|\tilde{x}\partial\tilde{x}x h_1|\tilde{0}\rangle\langle\tilde{0}|h_1^\prime x|\phi\rangle\langle\phi|x\bar{x}\partial\bar{x}h_2|\tilde{0}\rangle\langle\tilde{0}|h_2^\prime x\bar{x}\partial\bar{x}\tilde{x}\partial\tilde{x}|\phi\rangle.
\end{eqnarray}
By commuting some terms we obtain:
\begin{eqnarray}
P_2=|H^\prime|^{-2}\sum_{\substack{h_1,h_2\in H^\prime\\ h_1^\prime,h_1^\prime\in H^\prime}}\sum_{\substack{x\in X^\prime\\ \tilde{x}\in X_A^\prime\\ \bar{x}\in X_B^\prime}}\langle\phi|\tilde{x}\partial\tilde{x}h_1|\tilde{0}\rangle\langle\tilde{0}|h_1^\prime|\phi\rangle\langle\phi|h_2|\tilde{0}\rangle\langle\tilde{0}|h_2^\prime \tilde{x}\partial\tilde{x}|\phi\rangle\\ \nonumber
\times(-1)^{x\cap h_1h_1^\prime h_2 h_2^\prime}(-1)^{\bar{x}\cap h_2 h_2^\prime} (-1)^{\partial\bar{x}\cap h_1 h_2^\prime}.
\end{eqnarray}
The story here is just like the $P_1$ part. Repeating the derivation we get
\begin{eqnarray}
\label{P2f}
P_2(\partial\tilde{g},\partial\bar{g})=\frac{2^{2N}|X_B^\prime|}{|X^\prime|}\sum_{\tilde{h}\in H_A^\prime}\sum_{\tilde{x}\in X_A^\prime}|\langle\phi|\tilde{x}\partial\tilde{x}(\partial\tilde{g})\tilde{h}|\phi\rangle|^2(-1)^{\partial\bar{x}(\partial\bar{g})\cap\tilde{h}}.
\end{eqnarray}
The $2^{2N}$ term comes form $|X^\prime|=2^N|H^\prime|$.
Combing Eqs.($\ref{P1f}, \ref{P2f}$) we finally get the purity formula:
\begin{eqnarray}\label{purity}
P=C_P\sum_{\partial\tilde{g}\in \partial G^\prime_{A}} \sum_{\substack{\tilde{g}\in G^\prime_A\\ \tilde{z}\in Z_A}}|\langle\psi|\tilde{g}\partial\tilde{g}\tilde{z}|\psi\rangle|^2\sum_{\substack{\tilde{h}\in H^\prime_A\\ \tilde{x} \in X^{\prime}_{A}}}|\langle\phi|\tilde{x}\partial\tilde{x}(\partial\tilde{g})\tilde{h}|\phi\rangle|^2
\sum_{\partial\bar{g}\in \partial G^\prime_{B}}(-1)^{\partial\bar{g}\partial\bar{x}(\partial\bar{g})\cap\tilde{z}\tilde{h}},
\end{eqnarray}
where  the coefficient $C_P=\frac{|G^\prime_B|}{|G^\prime|}\frac{|X^{\prime}_{B}|}{|X^{\prime}|}$ can be presented as $2^{R+2}2^{-(\#A+\#\partial A)}|X^\prime_A|^{(-1)}$. Here, $\#A$ is the number of the site belonging to A and $\#\partial A$ is the number of site belonging to the boundary of the A and B. One can verify that the coefficient $C_P$ is vanished when we calculate the topological R\'enyi entropy.
According to Eq.($\ref{phase2}$), the last phase term $\sum_{\partial\bar{g}\in \partial G^\prime_{B}}(-1)^{\partial\bar{g}\partial\bar{x}(\partial\bar{g})\cap\tilde{z}\tilde{h}}$ selects some particular $\tilde{z}\tilde{h}$ out. They satisfy the condition $[\partial\bar{g}\partial\bar{x}(\partial\bar{g}), \tilde{z}\tilde{h}]=0$
 for all the $\partial\bar{g}\in \partial G^\prime_{B}$.

Exploiting the fact that
in `$\tau$-picture',  the state $|\psi\rangle$ , $|\phi\rangle$ and the operators occurred in Eq.($\ref{purity}$) have the tensor product form, we can decompose the operators and the state as the product form of each row and write the purity Eq.($\ref{purity}$) in the following form:
\begin{eqnarray}
P&=&C_P\prod_{k=1}^{R+2}\sum_{\partial\tilde{g}_{2k-1}\in\partial G^\prime_{A_{2k-1}}}
\sum_{\tilde{z}_{2k-1}\in Z^\prime_{2k-1}}P_{2k-1}(\partial\tilde{g}_{2k-1},\tilde{z}_{2k-1})\sum_{\tilde{h}_{2k}\in H^\prime_{2k}}P_{2k}\left(\partial\tilde{x}_{2k}
(\partial\tilde{g}_{2k-1},\partial\tilde{g}_{2k+1}),\tilde{h}_{2k}\right)\nonumber\\
&\times&\sum_{\partial\bar{g}_{2k-1}\in\partial G^\prime_{B_{2k-1}}}
(-1)^{\partial\bar{g}_{2k-1}\partial\bar{x}_{2k}(\partial\bar{g}_{2k-1},\partial\bar{g}_{2k+1})\cap\tilde{z}_{2k-1}\tilde{h}_{2k}}
\end{eqnarray}
where
\begin{eqnarray}\label{p2k-1}
P_{2k-1}(\partial\tilde{g}_{2k-1},\tilde{z}_{2k-1})=\sum_{\tilde{g}_{2k-1}\in G^\prime_{A_{2k-1}}}
|\langle\psi_{2k-1}|\tilde{g}_{2k-1}\partial\tilde{g}_{2k-1}\tilde{z}_{2k-1}|\psi_{2k-1}\rangle|^2,\\\label{p2k}
P_{2k}\left(\partial\tilde{x}_{2k}(\partial\tilde{g}_{2k-1},\partial\tilde{g}_{2k+1}),\tilde{h}_{2k}\right)
=\sum_{\tilde{x}_{2k}\in X^\prime_{A_{2k}}}|\langle\phi_{2k}|
\tilde{x}_{2k}\partial\tilde{x}_{2k}(\partial\tilde{g}_{2k-1},\partial\tilde{g}_{2k+1})\tilde{h}_{2k}|\phi_{2k}\rangle|^2.
\end{eqnarray}
We illustrate how to proceed with the   calculation by the case of subsystem (1). We start from k=1. The component of $\partial G^\prime_{A}$ on first row are $\mathbbm{1}_{1}$ while $\partial G^\prime_{B}$ is $\partial G^\prime_{B_1}$. Omitting the constant coefficient, the k=1 component of the purity is
\begin{eqnarray}\label{Pk1}
P(k=1)=
\sum_{\tilde{z}_{1}\in Z^\prime_{1}}P_{1}(\tilde{z}_{1})\sum_{\tilde{h}_{2}\in H^\prime_{A_{2}}}P_{2}(\tilde{h}_{2})
\sum_{\partial\bar{g}_{1}\in\partial G^\prime_{B_{1}}}
(-1)^{\partial\bar{g}_{1}\partial\bar{x}_{2}(\partial\bar{g}_{1})\cap\tilde{z}_{1}\tilde{h}_{2}}.
\end{eqnarray}
The constraint of $[\partial\bar{g}_1\partial\bar{x}_{2}(\partial\bar{g}_{1}), \tilde{z}_1\tilde{h}_2]=0$ directly give  that the summation of  $\sum_{\tilde{z}_{1}\in Z^\prime_{1}}$ and $\sum_{\tilde{h}_{2}\in H^\prime_{A_{2}}}$ are not independent. Notice that $\partial\bar{g}_1\partial\bar{x}_{2}(\partial\bar{g}_{1})$ constitute a group and it is generated by $\tau^z_{s^1_j}\tau^{x}_{s^2_{j}}\tau^{x}_{s^2_{j+1}}$ in `$\tau$-picture'. The fact that $[\tau^z_{s^1_j}\tau^{x}_{s^2_{j}}\tau^{x}_{s^2_{j+1}}, \tau^x_{s^1_j}\tau^x_{s^1_{j+1}}\tau^z_{s^2_{j}}]=0$ tell us that  $\tau^x_{s^1_j}\tau^x_{s^1_{j+1}}$ and $\tau^z_{s^2_{j}}$ always appear in the same time. Thus, $\tilde{h}_{2}$ is the function of $\tilde{z}_{1}$ and Eq.($\ref{Pk1}$) is
\begin{eqnarray}
P(k=1)=
\sum_{\tilde{z}_{1}\in Z^\prime_{1}}P_{1}(\tilde{z}_{1})P_{2}(\tilde{h}_{2}(\tilde{z}_{1})).
\end{eqnarray}

Next, we consider $ k=2,\cdots, r+1$ case in which $\partial G^{\prime}_{A_{2k-1}}$ and $\partial G^{\prime}_{B_{2k-1}}$ contain only identity $\mathbbm{1}_{2k-1}$ so the component of purity for this part is
\begin{eqnarray}
P(k=2,\cdots, r+1)=\prod_{k=2}^{r}\left[\sum_{\tilde{z}_{2k-1}\in Z^\prime_{2k-1}}P_{2k-1}(\tilde{z}_{2k-1})\right]\left[\sum_{\tilde{h}_{2k}\in H^\prime_{A_{2k}}}P_{2k}(\tilde{h}_{2k})\right]P_{2r+1}(\tilde{z}_{2r+1}).
\end{eqnarray}
It is clear that every row is independent with each other.
\begin{figure}[t]
\begin{center}
\includegraphics[width=0.5\textwidth]{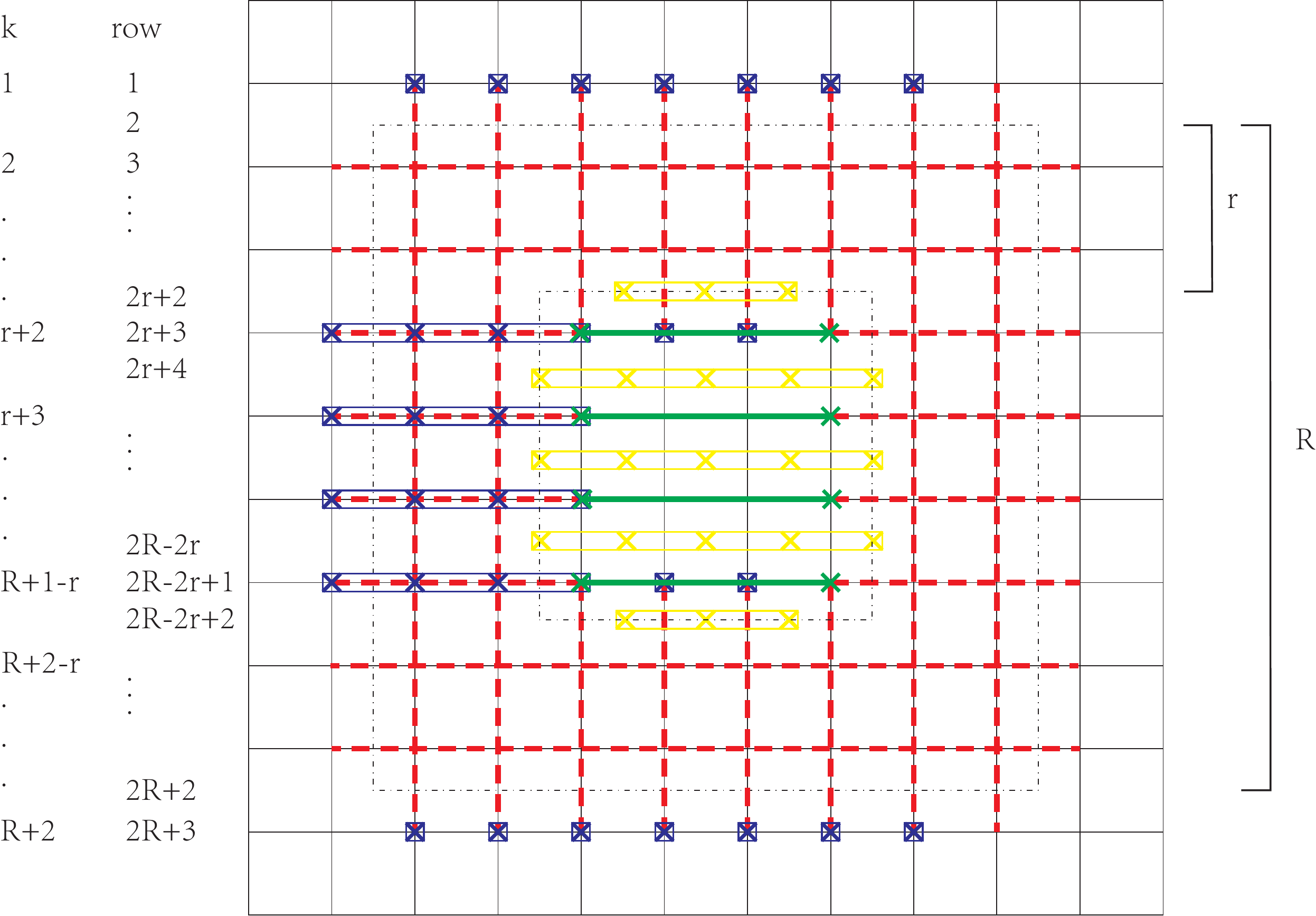}\\
\caption{(color online) Illustration of $\tilde{h}_{2k}^c$ (yellow rectangle block with cross), $\tilde{z}_{2k-1}^{c}$ (green solid line with cross ends) in subsystem (1) with R=8,r=2.}\label{connectoperator}
\end{center}
\end{figure}
Our aim is to get the topological R\'enyi entropy $S^T_2=log_2\left({P^{(1)}P^{(4)}}/P^{(2)}P^{(3)}\right)$ and we can finally find that $P(k=1,\cdots,r+1)=P(k=1)P(k=2,\cdots,r+1)$ for subsystem (1) is canceled during the calculation. The part that we really concern is the rows which contain the `hole'. We can write the purity of subsystem (1) as $ P^{(1)}=P_{\text{top}}P_{\text{hole}}^{(1)}P_{\text{bottom}}$ where $P_{\text{top}}=P_{\text{bottom}}=P(k=1,\cdots,r+1)$ ( caused by the symetry of the subsystem and the fields ). For subsystem (2) and (3) they are $ P^{(2)}=P_{\text{top}}P_{\text{hole}}^{(2)}$  and $ P^{(3)}=P_{\text{hole}}^{(2)}P_{\text{bottom}}$. For subsystem (4) , every row is disconnected so $S^T_2=log_2\left(P^{(1)}_{\text{hole}}P^{(4)}/P^{(2)}_{\text{hole}}P^{(3)}_{\text{hole}}\right)$. We take subsystem (1) for example to calculate $P^{(1)}_{\text{hole}}$:
\begin{eqnarray}
P_{\text{hole}}^{(1)}&=&\prod_{k=r+2}^{R+1-r}\sum_{\partial\tilde{g}_{2k-1}}P_{2r+2}\left(\partial\tilde{x}_{2r+2}
(\partial\tilde{g}_{2r+3}),\tilde{h}_{2r+2}\right)\nonumber\\
&\times&\sum_{\tilde{z}_{2k-1}}P_{2k-1}(\partial\tilde{g}_{2k-1},\tilde{z}_{2k-1})\sum_{\tilde{h}_{2k}}P_{2k}\left(\partial\tilde{x}_{2k}
(\partial\tilde{g}_{2k-1},\partial\tilde{g}_{2k+1}),\tilde{h}_{2k}\right)\nonumber\\
&\times&\sum_{\prod_{k=r+2}^{R+1-r}\partial\bar{g}_{2k-1}}
(-1)^{\partial\bar{x}_{2r+2}\prod_{k=r+2}^{R+1-r}\partial\bar{g}_{2k-1}\partial\bar{x}_{2k}\cap\tilde{h}_{2r+2}\prod_{k=r+2}^{R+1-r}\tilde{z}_{2k-1}\tilde{h}_{2k}}.
\end{eqnarray}
We use the the shorthand notation to denote the last phase term
\begin{eqnarray}
\sum_{\partial\bar{g}_{h}}
(-1)^{\partial\bar{g}_{h}\partial\bar{x}(\bar{g}_{m})\cap\tilde{z}_{m}\tilde{h}_{m}}.
\end{eqnarray}
It gives a constrain in order to fulfil this term being nonzero. showing as  Fig.\ref{partialGAGB} and Fig.\ref{connectoperator}, the constrain is caused by $\partial\bar{g}_h$  of type (ii) and (iii). We denote $\prod_{j=1}^{r+2}\tau^z_{s^{2k-1}_j}$ as $\partial\bar{g}^{c}_{2k-1}$ (blue rectangle block), $\prod_{j=r+2}^{R-r}\tau^z_{s^{2k}_j}$ as $\tilde{h}_{2k}^c$ (yellow rectangle block) and $\tau^x_{s^{2k-1}_{r+2}}\tau^x_{s^{2k-1}_{R+1-r}}$ as $\tilde{z}_{2k-1}^{c}$ (green solid line with cross ends). Notice the following relationship: $\{\partial\bar{g}^{c}_{2k-1} , \tilde{z}_{2k-1}^{c}\}=0$, $\{\partial\bar{x}(\partial\bar{g}^{c}_{2k-1}) , \tilde{h}_{2k-2}^c\}=0$ and $\{\partial\bar{x}(\partial\bar{g}^{c}_{2k-1}) , \tilde{h}_{2k}^c\}=0$. We can get that for $k=(r+2),\cdots, (R+1-r)$, every $(\tilde{h}_{2k-2}^c)^{m_{2k-2}}(\tilde{z}_{2k-1}^{c})^{m_{2k-1}}(\tilde{h}_{2k}^c)^{m_{2k}}$
should obey ($m_{2k-2}+m_{2k-1}+m_{2k}$)mod2=0 (\{m\}=0,1). The number of possible configuration is $2^{(R-2r+1)}$. We choose R=5, r=1, for example, then the number is 16. every Column in the Table \ref{mconfig} represents one possible configuration \{m\}.
\begin{table}
	\begin{tabular}{|c|l|l|l|l|l|l|l|l|l|l|l|l|l|l|l|l|}
	$m_4$ & 0 & 0 & 0 & 0 & 0 & 0 & 0 & 0 & 1 & 1 & 1 & 1 & 1 & 1 & 1 & 1 \\
	$m_5$ & 0 & 0 & 0 & 0 & 1 & 1 & 1 & 1 & 0 & 0 & 0 & 0 & 1 & 1 & 1 & 1 \\
	$m_6$ & 0 & 0 & 0 & 0 & 1 & 1 & 1 & 1 & 1 & 1 & 1 & 1 & 0 & 0 & 0 & 0 \\
	$m_7$ & 0 & 0 & 1 & 1 & 0 & 0 & 1 & 1 & 0 & 0 & 1 & 1 & 0 & 0 & 1 & 1 \\
	$m_8$ & 0 & 0 & 1 & 1 & 1 & 1 & 0 & 0 & 1 & 1 & 0 & 0 & 0 & 0 & 1 & 1 \\
	$m_9$ & 0 & 1 & 0 & 1 & 0 & 1 & 0 & 1 & 0 & 1 & 0 & 1 & 0 & 1 & 0 & 1 \\
	$m_{10}$ & 0 & 1 & 1 & 0 & 1 & 0 & 0 & 1 & 1 & 0 & 0 & 1 & 0 & 1 & 1 & 0 \\
	\end{tabular}
\caption{ All possible configurations of \{m\} obeying ($m_{2k-2}+m_{2k-1}+m_{2k}$)mod2=0 for $k=(r+2),\cdots, (R+1-r)$ with subsystem size R=5,r=1.}\label{mconfig}
\end{table}

We denote $\sum_{\{\tilde{z},\tilde{h}\}_{\text{constr.}}}$ as the summation of $\tilde{z},\tilde{h}$ in all the possible configurations, which satisfy the constrain.
Finally we get
\begin{eqnarray}
P^{(1)}_{\text{hole}}&=&C^{(1)}_{\text{hole}}\prod_{k=r+2}^{R+1-r}\sum_{\partial\tilde{g}_{2k-1}}\sum_{\{\tilde{z},\tilde{h}\}_{\text{constr.}}}P_{2r+2}\left(\partial\tilde{x}_{2r+2}
(\partial\tilde{g}_{2r+3}),\tilde{h}_{2r+2}\right)\nonumber\\
&\times&P_{2k-1}(\partial\tilde{g}_{2k-1},\tilde{z}_{2k-1})P_{2k}\left(\partial\tilde{x}_{2k}
(\partial\tilde{g}_{2k-1},\partial\tilde{g}_{2k+1}),\tilde{h}_{2k}\right).
\end{eqnarray}
The coefficient comes from Eq.(\ref{phase2}) and it is vanished when we calculate the topological R\'enyi entropy. Together with Eq.($\ref{p2k-1}$, $\ref{p2k}$),  the final task is to calculate the square expectation values of form $|\langle\psi|\tau^z_{m_1}\cdots\tau^z_{m_s}\cdots\tau^x_{n_1}\cdots\tau^x_{n_{2t}}|\psi\rangle|^2$ each rows.
We concern the state $|\psi\rangle$ in two conditions. (i) The static one which is the ground state of the Ising Hamiltonian showed in Eq.($\ref{isingH})$ each row. (ii) Time evolution state after a quantum quench . Both of them can be treated analytically in free fermion representation.
\end{widetext}

\subsection{Calculation of purity: mapping to free fermion }
As we saw in the previous section, the purity $P$ can then be directly calculated in terms of expectation values of strings of $\tau$ operators. These expectation values can be computed exactly in the case of the integrable chain\cite{barouch:1971, isingbook}. We are concerned with
one dimensional Ising model in transverse field:
\begin{eqnarray}
H_{\text{Ising}}=-\sum_{l=1}^{N}(\tau^{z}_l+\lambda\tau_l^x\tau_{l+1}^x).
\end{eqnarray}
We take the standard procedure of Jordan-Wigner transformation to map the Hamiltonian to the free fermion representation.
\begin{eqnarray}
\sigma_l^z=1-2c^{\dagger}_l c_l,\quad \sigma_l^{+}=(\sigma_l^{-})^\dagger=\prod_{j=1}
^{l-1} (1-2c^{\dagger}_j c_j)c_l.\nonumber\\
\end{eqnarray}
Note that we map the spin up state $|\uparrow\rangle$ to the vacuum $|0\rangle$. So the spin Hamiltonian transforms to the fermion Hamiltonian
\begin{eqnarray}
H_{Ising}=&&-\sum_{l=1}^N(c_l^\dagger+c_l)(c_l^\dagger-c_l)\nonumber\\
&&-\lambda\sum_{l=1}^N (c_l^\dagger-c_l)(c_{l+1}^\dagger+c_{l+1})\nonumber\\
&&+\lambda[exp(i\pi\sum_{j=1}^n c_j^\dagger c_j)+1](c_N^\dagger-c_N)(c_1^\dagger+c_1).\nonumber\\
\end{eqnarray}
Notice that the operator identity $1-2c_j^\dagger c_j=(c_j^\dagger+c_j)(c_j^\dagger-c_j)=exp(i\pi c_j^\dagger c_j)$ and the term $exp(i\pi\sum_{j=1}^n c_j^\dagger c_j)$ is actually the parity operator of the system which is 1 in the subspace we choose. Nevertheless, the last term is the correction term which we neglect in the large N limit.
 As the former formula showed, what we are interested in is some square expectation value in shape of $|\langle\psi|\tau^z_{m_1}\cdots\tau^z_{m_s}\cdots\tau^x_{n_1}\cdots\tau^x_{n_{2t}}|\psi\rangle|^2$. Defining $A_j=c_j^\dagger+c_j$, $B_j=c_j^\dagger-c_j$ ,
it's easy to check that $\sigma_j^z=A_j B_j$ and $\sigma_j^x\sigma_{j+1}^x=B_j A_{j+1}$. Noting that $A_j^2=1$ and $B_j^2=-1$, the former equation can be written in the form of $|\langle\psi| \cdots A_s \cdots B_t \cdots|\psi\rangle|^2$. Wick's theorem tells us this expression can be reduced to the product of two-operator expectation values.

Applying Fourier transformation
\begin{eqnarray}
c_l=\frac{1}{\sqrt{N}}\sum_q e^{iql} c_q,
\end{eqnarray}
the Hamiltonian is rewritten in  momentum space as
\begin{eqnarray}
H=&&\sum_q(1-\lambda \cos q)(c_q^\dagger c_q-c_{-q}c_{-q}^\dagger)\nonumber\\
&&-\lambda i\sum_q \sin q(c_q^\dagger c_{-q}^\dagger-c_{-q}c_q)\nonumber \\
=&& \sum_q C_q^\dagger M_q(\lambda)C_q,
\end{eqnarray}
where
\begin{eqnarray}
M_q(\lambda)=\left(
               \begin{array}{cc}
                 a_q(\lambda) & -ib_q(\lambda) \\
                 ib_q(\lambda) & -a_q(\lambda) \\
               \end{array}
             \right),\nonumber\\
             \quad a_q(\lambda)=1-\lambda \cos q, \quad b_q(\lambda)=\lambda \sin q,
\end{eqnarray}
and
\begin{eqnarray}
C_q=\left(
      \begin{array}{c}
        c_q \\
        c_{-q}^\dagger \\
      \end{array}
    \right).
\end{eqnarray}
The Hamiltonian is diagonalised by the Bogoliubov transformation:
\begin{eqnarray}
H=\sum_q h_q(\lambda)^\dagger\left(
                               \begin{array}{cc}
                                 \omega_q(\lambda) & 0 \\
                                 0 &-\omega_q(\lambda)  \\
                               \end{array}
                             \right)h_q(\lambda)
\end{eqnarray}
where
\begin{eqnarray}\label{Rmatrix}
h_q(\lambda)&=&\left(
               \begin{array}{c}
                 \eta_q(\lambda) \\
                 \eta_{-q}^\dagger(\lambda) \\
               \end{array}
             \right)=R_q^\dagger(\lambda)C_q,\nonumber\\
              R_q(\lambda)&=&\left(
                                                                  \begin{array}{cc}
                                                                    u_q(\lambda) & -iv_q(\lambda) \\
                                                                    -iv_q(\lambda) & u_q(\lambda) \\
                                                                  \end{array}
                                                                \right)
\end{eqnarray}
with
\begin{eqnarray}
u_q(\lambda)&=&\frac{a_q(\lambda)+\omega_q(\lambda)}{\sqrt{2\omega_q(\lambda)(\omega_q(\lambda)+a_q(\lambda))}},\nonumber\\
v_q(\lambda)&=&\frac{-b_q(\lambda)}{\sqrt{2\omega_q(\lambda)(\omega_q(\lambda)+a_q(\lambda))}}
\end{eqnarray}
and
\begin{eqnarray}
\omega_q(\lambda)=\sqrt{a_q(\lambda)^2+b_q(\lambda)^2}=\sqrt{1-2\lambda cosq+\lambda^2}.\nonumber\\
\end{eqnarray}

After diagonalising the Hamiltonian, we can obtain the exact eigenstates. Moreover, we can obtain an exact expression for the time evolution. In the quantum quench scenario, the state evolves as
\begin{eqnarray}
|\Psi(t)\rangle=U(t)|\Psi(0)\rangle=e^{-itH(\lambda(t))}|\Psi(0)\rangle.
\end{eqnarray}
in which
\begin{eqnarray}\label{lambdat}
\lambda(t)=\left\{
\begin{aligned}
&\lambda_0\quad&(t\leqslant0)\\
&\lambda\quad&(t>0)
\end{aligned}
\right.
\end{eqnarray}
and the initial state $|\Psi(0)\rangle$ is the ground state of $H(\lambda_0)$, namely $\eta_q(\lambda_0)|\Psi(0)\rangle=0$, $\forall q$. (Basically what we are interested in is the condition of $\lambda_0=0$ in which the following derivation would be simplified, but we do the derivation in general condition.)
We need to calculate the expectation value of $\langle\Psi(t)|O|\Psi(t)\rangle=\langle\Psi(0)|e^{itH(\lambda(t))}Oe^{-itH(\lambda(t))}|\Psi(0)\rangle=\langle\Psi(0)|O^H(t)|\Psi(0)\rangle$, and operator O we concerning is the product of some $A_j$ and $B_k$ operators. So we need to applying the Wick's theorem in Heisenberg picture.
First we focus on $c_q^H(t)$ whose Heisenberg equation is
\begin{eqnarray}
i\frac{d}{dt}c_q^H(t)&=&U^\dagger(t)[c_q, H(\lambda(t))]U(t)\nonumber\\
&=&2a_q(\lambda(t))c_q^H(t)-2ib_q(\lambda(t))c_{-q}^H(t)^\dagger\nonumber\\
\end{eqnarray}
or more compactly
\begin{eqnarray}
i\frac{d}{dt}C_q^H(t)=2M_q(\lambda(t))C_q^H(t).
\end{eqnarray}
Expand $C_q^H(t)$ by $\eta_q(\lambda_0)$ and $\eta_{-q}^\dagger(\lambda_0)$ as $C_q^H(t)=S_q(t)h_q(\lambda_0)$, thus
\begin{eqnarray}
S_q(t)=\left(
         \begin{array}{cc}
           \tilde{u}_q(t) & -\tilde{v}_q^\ast(t) \\
           \tilde{v}_q(t) & \tilde{u}_q^\ast(t) \\
         \end{array}
       \right)
\end{eqnarray}
which is unitary and obeys the constrains
\begin{eqnarray}
\tilde{u}_q(t)=\tilde{u}_{-q}(t), \tilde{v}_q(t)=-\tilde{v}_{-q}(t).
\end{eqnarray}
Then we obtain equations of motion for  $S_q(t)$:
\begin{eqnarray}
i\frac{d}{dt}\left(
               \begin{array}{c}
                 \tilde{u}_q(t) \\
                 \tilde{v}_q(t) \\
               \end{array}
             \right)=2\left(
               \begin{array}{cc}
                 a_q(\lambda(t)) & -ib_q(\lambda(t)) \\
                 ib_q(\lambda(t)) & -a_q(\lambda(t)) \\
               \end{array}
             \right)\left(
               \begin{array}{c}
                 \tilde{u}_q(t) \\
                 \tilde{v}_q(t) \\
               \end{array}
             \right).\nonumber\\
\end{eqnarray}
Combining Eqs.(\ref{Rmatrix}) and (\ref{lambdat}) the solution is
\begin{widetext}
\begin{eqnarray}
\left(
               \begin{array}{c}
                 \tilde{u}_q(t) \\
                 \tilde{v}_q(t) \\
               \end{array}
             \right)=R_q(\lambda)\left(
                                   \begin{array}{cc}
                                     e^{-i2\omega_q(\lambda)t} & 0 \\
                                     0 & e^{i2\omega_q(\lambda)t} \\
                                   \end{array}
                                 \right)R_q^\dagger(\lambda)\left(
               \begin{array}{c}
                 \tilde{u}_q(0) \\
                 \tilde{v}_q(0) \\
               \end{array}
             \right).
\end{eqnarray}
\end{widetext}
Combining with the initial condition
\begin{eqnarray}
\left(
               \begin{array}{c}
                 \tilde{u}_q(0) \\
                 \tilde{v}_q(0) \\
               \end{array}
               \right)=\left(
               \begin{array}{c}
                 u_q(\lambda_0) \\
                 -iv_q(\lambda_0) \\
               \end{array}
               \right)
\end{eqnarray}
we finally get the solution:
\begin{eqnarray}\label{uvsolution}
\left(
               \begin{array}{c}
                 \tilde{u}_q(t) \\
                 \tilde{v}_q(t) \\
               \end{array}
             \right)=\left(
                       \begin{array}{c}
                         u_0cos2\omega t+i(\frac{-au_0+bv_0}{\omega})sin2\omega t \\
                         -iv_0cos2\omega t+(\frac{bu_0+av_0}{\omega})sin2\omega t  \\
                       \end{array}
                     \right)\nonumber\\
\end{eqnarray}
where we have applied the shorthand notations: $a=a_q(\lambda), b=b_q(\lambda), \omega=\omega_q(\lambda), u_0=u_q(\lambda_0)$, and $ v_0=v_q(\lambda_0)$.

{ It's the time to work on the operators $A_j^H(t)$ and $B_j^H(t)$. As the standard procedure of applying Wick's theorem, we need to decompose the operators in two parts: $A_j^H(t)=a_j^\dagger(t)+a_j(t)$ and $B_j^H(t)=b_j^\dagger(t)-b_j(t)$, where
\begin{eqnarray}\label{ab}
a_j(t)=\frac{1}{\sqrt{N}}\sum_q e^{iqj}(\tilde{u}_q(t)+\tilde{v}_q(t))\eta_q(\lambda_0)\nonumber\\
b_j(t)=\frac{1}{\sqrt{N}}\sum_q e^{iqj}(\tilde{u}_q(t)-\tilde{v}_q(t))\eta_q(\lambda_0).
\end{eqnarray}
Notice that $a_j(t)$ (also $b_j(t)$) is a combination of $\eta_q(\lambda_0)$, which are the destruction operators acting on the initial state $|\Psi(0)\rangle$, while $a_j^\dagger(t)$ (also $b^\dagger_j(t)$) is a combination of $\eta^\dagger_q(\lambda_0)$.

Actually $\{\eta_q(\lambda_0)\}$ and $\{\eta^\dagger_q(\lambda_0)\}$ form a set of bases of the operators in the Hilbert-space, so any operator $O$ have a decomposition $O=O^-+O^+$, where $O^-|\Psi(0)\rangle$=0 and $\langle\Psi(0)|O^+=0$. Two operators product can be written as $O_1O_2=N[O_1O_2]+\{A_1^-,A_2^+\}$, where N is the normal ordering operator and the anti-commutator comes from the fermi statistics. So the expectation value satisfies $\langle\Psi(0)|O_1O_2|\Psi(0)\rangle=\{A_1^-,A_2^+\}$, which is known as contraction of two operators.

Following three types of contraction are concerned:
\begin{eqnarray}
G_{j-k}(t)&=&\langle\Psi(0)|A_j^H(t)B_k^H(t)|\Psi(0)\rangle=\{a_j(t),b_k^\dagger(t)\},\nonumber\\
G_{j-k}^A(t)&=&\langle\Psi(0)|A_j^H(t)A_k^H(t)|\Psi(0)\rangle=\{a_j(t),a_k^\dagger(t)\},\nonumber\\
G_{j-k}^B(t)&=&\langle\Psi(0)|B_j^H(t)B_k^H(t)|\Psi(0)\rangle=-\{b_j(t),b_k^\dagger(t)\}.\nonumber\\
\end{eqnarray}
Substituting (\ref{ab}), They can be written explicitly as
\begin{eqnarray}
G_{j-k}(t)&=&\frac{1}{N}\sum_qe^{iq(j-k)}(|\tilde{u}_q(t)|^2-|\tilde{v}_q(t)|^2\nonumber\\
&&+\tilde{v}_q(t)\tilde{u}_q^\ast(t)-\tilde{u}_q(t)\tilde{v}_q^\ast(t)),\nonumber\\
G_{j-k}^A(t)&=&\delta_{j,k}+\frac{1}{N}\sum_qe^{iq(j-k)}(\tilde{v}_q(t)\tilde{u}_q^\ast(t)\nonumber\\
&&+\tilde{u}_q(t)\tilde{v}_q^\ast(t)),\nonumber\\
G_{j-k}^B(t)&=&-\delta_{j,k}+\frac{1}{N}\sum_qe^{iq(j-k)}(\tilde{v}_q(t)\tilde{u}_q^\ast(t)\nonumber\\
&&+\tilde{u}_q(t)\tilde{v}_q^\ast(t)).
\end{eqnarray}
Noticing that $G_{j-k}^\ast(t)=G_{j-k}(t)$, so the contraction $\langle B_j^H(t)A_k^H(t)\rangle=-\langle A_k^H(t)B_j^H(t)\rangle^\ast=-G_{-(j-k)}$.

$\tilde{u}_q(t)$ and $\tilde{v}_q(t)$ are solved showed in Eq.(\ref{uvsolution}), so we can get the final contraction formulas:
\begin{eqnarray}
G_{r}(t)&=&\frac{1}{N}\sum_qe^{iqr}\left(\frac{a_0a+b_0b}{(a-ib)\omega_0}+icos4\omega t\frac{ab_0-a_0b}{(a-ib)\omega_0}\right), \nonumber\\
G_{r}^A(t)&=&\delta_{r,0}+\frac{1}{N}\sum_qe^{iqr}\left(\frac{-ab_0+a_0b}{\omega\omega_0}\right)sin4\omega t, \nonumber\\
G_{r}^B(t)&=&-\delta_{r,0}+\frac{1}{N}\sum_qe^{iqr}\left(\frac{-ab_0+a_0b}{\omega\omega_0}\right)sin4\omega t.
\end{eqnarray}

These formulas of contraction are derived from the quantum quench scenario. However, if the quench Hamiltonian stays constant, that is $\lambda(t)=\lambda_0$, the initial state will not evolve. We will now we discuss the formula in both two cases.

\textbf{(1) static case}

In this case $\lambda_0=\lambda$. So we have $a_0=a=a_q(\lambda)=1-\lambda cosq$, $b_0=b=b_q(\lambda)=\lambda sinq$ and $\omega_0=\omega=\omega_q(\lambda)=\sqrt{1-2\lambda cosq+\lambda^2}$. It gives the contraction formulas in static case directly:
\begin{eqnarray}
G_r&=&\frac{1}{N}\sum_qe^{iqr}\left(\frac{a-ib}{\omega}\right),\nonumber\\
G_r^A&=&\delta_{r,0},~~
G_r^B=-\delta_{r,0}.
\end{eqnarray}
The time-dependent term vanish automatically during the derivation in this condition.

\textbf{(2) quantum quench (time dependent) case}

Basically we concern the initial state is the ground state of the toric code model, which corresponds to $\lambda_0=0$. In this case we have $a_0=1$, $b_0=0$ and $\omega_0=1$. We concern long-time evolution in this paper so the limit $t\rightarrow\infty$ is reasonable. Taking thermodynamic limit $\frac{1}{N}\rightarrow\int dq/2\pi$, the time-dependent term vanishes caused by the fast oscillation (Lebesgue lemma). After an algebra, We can get
\begin{eqnarray}
G_r(\infty)=\frac{1}{2\pi}\int_{-\pi}^{\pi}dq e^{-iqr}\frac{a}{(a+ib)}.
\end{eqnarray}
Remembering that $a=a_q(\lambda)=1-\lambda cosq$ and $b=b_q(\lambda)=\lambda sinq$, so}
\begin{eqnarray}
G_r(\infty)=\frac{1}{2\pi}\int_{-\pi}^{\pi}dq e^{-iqr}\frac{2-\lambda(e^{iq}+e^{-iq})}{2(1-\lambda e^{-iq})}.
\end{eqnarray}
Changing the integral to an contour integral on the complex plane of $z=e^{iq}$:
\begin{eqnarray}
G_r(\infty)=\frac{1}{2\pi i}\oint dz z^{-r-1}\frac{-\lambda z^2+2z-\lambda}{2(z-\lambda)}
\end{eqnarray}
where the integral path is along the unit circle. Applying the residue theorem we can finally obtain the exact value of $G_r(\infty)$. For $\lambda<1$,
\begin{eqnarray}
G_r(\infty)=\left\{
\begin{aligned}
    &0\quad&(&r\geq2)\\
    &-\lambda/2\quad&(&r=1)\\
    &-\frac{1}{2}\lambda^2+1\quad&(&r=0)\\
    &\frac{1}{2}(\lambda)^{-r}(1-\lambda^2)\quad&(&r\leq-1)
\end{aligned}
\right.,
\end{eqnarray}
and for $\lambda>1$,
\begin{eqnarray}
G_r(\infty)=\left\{
\begin{aligned}
    &\frac{1}{2}\lambda^{-r}(\lambda^2-1)\quad&(&r\geq2)\\
    &-\frac{1}{2\lambda}\quad&(&r=1)\\
    &\frac{1}{2}\quad&(&r=0)\\
    &0\quad&(&r\leq-1)
\end{aligned}
\right..
\end{eqnarray}
We can also get $G_r^A(\infty)=\delta_{r,0}$ and $G_r^B(\infty)=-\delta_{r,0}$ thus all types of contraction are known. The expectation value of $|\langle\Psi(t)| \ldots A_s \ldots B_t \ldots|\Psi(t)\rangle|^2$ can be then directly calculated.\\

%

\end{document}